\RequirePackage{fix-cm}
\documentclass[smallextended]{svjour3}       % onecolumn (second format)
\smartqed  % flush right qed marks, e.g. at end of proof
\usepackage[top=2cm, bottom=2.5cm, left=2cm, right=2cm]{geometry}
\usepackage{pythonhighlight}
\usepackage{blkarray}\usepackage{amsmath}
\usepackage{amssymb}

\DeclareMathOperator*{\argmin}{arg\,min}

\usepackage{multirow}
\usepackage{makecell}
\usepackage{graphicx}
\usepackage{color}%��ɫ
\usepackage[justification=centering]{caption}

\usepackage{tabularx}
\usepackage{dcolumn}
\usepackage{amsmath}
\usepackage{amssymb}
\usepackage{booktabs}
\usepackage{subfigure}
\usepackage{float}
 \usepackage{bm}
\usepackage[mathscr]{euscript}
\usepackage[ruled,linesnumbered]{algorithm2e}
\begin{document}

\title{Analysis of uncertain fixed-effects model for  Latin square designs%\thanks{Grants or other notes
%about the article that should go on the front page should be
%placed here. General acknowledgments should be placed at the end of the article.}
}
\subtitle{}

%\titlerunning{Short form of title}        % if too long for running head

\author{Yaru Cheng  \and
        Zhiming Li*%etc.
}

%\authorrunning{Short form of author list} % if too long for running head

\institute{ College of Mathematics and System Science, Xinjiang University,  Urumqi 830046, China \\
             % Tel.: +15276759658\\
              \email{zmli@xju.edu.cn}           %  \\
}

\date{Received: date / Accepted: date}
% The correct dates will be entered by the editor

\maketitle

\begin{abstract}
Uncertain data without frequency stability often arises in experimental design.  Classical fixed-effects models can only analyze precise experimental data.   Based on an uncertain measure, this paper establishes uncertain fixed-effect models for Latin-square designs.  First, we propose three methods with uncertainty to estimate the treatment and blocked effects and construct their confidence intervals. Then, uncertain homogeneity and common tests are conducted to assess the significance of treatment effects. In the numerical simulations, the three estimation methods are compared based on bias, mean squared error, mean absolute error, overall standard deviation, coverage probability, and average interval length.
Several examples are given to illustrate the process of estimation and hypothesis. Finally, the uncertain fixed-effects model is applied to real education data, demonstrating its practical value.
	
% \PACS{PACS code1 \and PACS code2 \and more}
% \subclass{MSC code1 \and MSC code2 \and more}
\end{abstract}

\section{Introduction}\label{s1}
Latin square designs, proposed by Fisher \cite{2}, have been widely applied across various fields such as medicine \cite{5}, information science \cite{3}, and computer science \cite{4}. It can effectively control and eliminate the influence of two orthogonal blocking factors on experimental outcomes.  Osuagwu et al. \cite{1} showed that Latin square designs can significantly outperform completely randomized designs in terms of statistical power and estimation accuracy under certain conditions. Current research on Latin square designs mainly focuses on two aspects: design construction \cite{7,8,9} and statistical modeling \cite{6}. Among these, statistical modeling-based Latin designs are the core component of experimental design and are crucial for valid inference.  In the probability measure framework, traditional modeling approaches for Latin square designs typically rely on additive models \cite{2}. Subsequent extensions have been developed within different frameworks: the model-based generalized Latin square (Tar \cite{10}), the Neyman model (Ogawa \cite{11}), and the general linear model (John \cite{12}, Onyiah \cite{13}). 

Although probability-based Latin square designs have developed into a well-established theoretical system, they still face substantial challenges when handling experimental data characterized by subjective uncertainty or incomplete information.  One of the main reasons is that the robustness of probabilistic models depends heavily on the strict assumptions of three axioms: non-negativity, normality, and countable additivity. In practice, some observed data often exhibit skewness and outliers, and are often based on subjective assessments with imprecise information; the validity of statistical inference may be severely compromised \cite{14}. To address these issues, researchers have proposed various extensions; for example, Aita and Talebi \cite{15} applied the augmented Latin square design to imprecise data, and Kumari et al. \cite{16} studied the neutrosophic Latin square analysis for uncertain data. To analyze the uncertainty of such data, Liu \cite{17} first proposed an uncertain measure satisfying four axioms: normality, duality, subadditivity, and product, providing an alternative mathematical framework \cite{18}. As a central application of the uncertainty theory, uncertainty statistics, particularly uncertain regression analysis, provide systematic tools for handling uncertain data \cite {19}. At present, uncertain regression has achieved considerable maturity in parameter estimation, encompassing methods such as maximum likelihood estimation \cite{20}, least absolute deviation estimation \cite{21}, and robust approaches based on the Huber loss and ridge regression \cite{22}. Recent developments in robust uncertain regression and its practical applications have also attracted considerable attention, for example, the least absolute deviations estimator has been applied to residual analysis and sports statistics to alleviate the adverse effects of outliers \cite{29,30}, the least relative error estimator has been adopted to characterize dynamic error structures, leading to enhanced fitting performance in the presence of anomalous observations \cite{31}, the uncertainty-theory-based Huber support vector regression (Huber-SVR) model \cite{32} and the robust kernel randomized configuration network \cite{33} have demonstrated satisfactory performance in handling imprecise observational data and solving complex industrial regression problems, respectively. In addition, Ye and Liu \cite{23,24} introduced hypothesis testing into uncertain regression analysis, enabling significance testing of pre-specified regression coefficients under uncertainty. 

Up to now, uncertainty statistics have become increasingly mature in the field of regression analysis. However,  the uncertain modeling-based Latin square designs remain relatively underexplored. Motivated by this gap, this paper aims to construct an uncertain fixed-effects model based on the Latin square design and to systematically derive the parameter estimation and inference methods. This research not only extends the analytical framework of experimental design in uncertain environments at a theoretical level but also provides robust statistical tools for handling uncertain data in fields such as agricultural experiments, engineering assessment, and management science. Accordingly, the main contributions of this study are summarized as follows:
	
(i) An uncertain fixed effects model based on the Latin square design is constructed, which is capable of characterizing uncertainty arising from subjective cognition and incomplete information.

(ii) A comprehensive comparison of multiple estimation methods is conducted via six criteria: bias, mean squared error, mean absolute error, overall standard deviation, coverage probability, and average interval length. Compared with a single evaluation metric, this six-criterion framework assesses performance from multiple perspectives, thereby achieving a better balance in the estimation results.

(iii) Homogeneity tests and common procedures are developed under both homoscedasticity and heteroscedasticity conditions. Compared with existing methods that are mainly applicable to multifactor settings, the proposed approach can effectively handle row and column blocking effects.

(iv) Numerical simulations are conducted to systematically validate the proposed model and methods, evaluating their estimation and testing performance under various scenarios. The uncertain fixed-effects model is applied to educational problems, demonstrating the practical applicability of  Latin square designs.

The rest of this paper is organized as follows. In Section~\ref{s2}, we review basic concepts of uncertainty theory and establish an uncertain fixed-effects model for Latin square designs. The parameter estimations and confidence interval methods are proposed in Section \ref{s3}. Section~\ref{s4} conducts the homogeneity and common tests to investigate the significance of treatment and block factors in Latin square designs. Section~\ref{s5} presents numerical simulation and a case study. Finally, a brief conclusion is given in Section~\ref{s6}.

%%%%%%%%%%%%%%%%%%%%%%%%%%%%%%%%%%%%%%%%%%%%%%%%%
\section{Preliminaries and Model Description}\label{s2}

  In this section, we first review basic definitions, such as the uncertain measure, the uncertain variable, and the uncertain distribution, from Liu \cite{17,18}, and then establish uncertain fixed-effects models for Latin square designs.

%%%%%%%%%%%%%%%%%%%%%%%%%%%%%%%%%%%%%%%%%%%%%%%%%
\subsection{Basic Concepts and Lemmas}
%%%%%%%%%%%%%%%%%%%%%%%%%%%%%%%%%%%%%%%%%%%%%

 Let $\Gamma$ be a nonempty set and $\mathcal{L}$ be a $\sigma$-algebra over $\Gamma$.  
A set function $\mathcal{M}: \mathcal{L} \rightarrow [0,1]$ is called an \textit{uncertainty measure} if it satisfies the following four axioms:

(i) \textbf{(Normality Axiom)}  
	For the universal set $\Gamma$,  $\mathcal{M}\{\Gamma\} = 1.	$
    
(ii) \textbf{(Duality Axiom)}  
	For any event $\Lambda\in \mathcal{L}$, $	\mathcal{M}\{\Lambda\} + \mathcal{M}\{\Lambda^{c}\} = 1.$
    
(iii) \textbf{(Subadditivity Axiom)}  
	For any countable sequence of events $\{\Lambda_i\}\subset \mathcal{L}$,  we have
	$$
	\mathcal{M}\left\{\bigcup_{i=1}^{\infty}\Lambda_i\right\} 
	\leq 
	\sum_{i=1}^{\infty}\mathcal{M}\{\Lambda_i\}.
	$$

(iv) \textbf{(Product Axiom)}  
	Let $(\Gamma_i,\mathcal{L}_i,\mathcal{M}_i)$ be a sequence of uncertainty spaces.  
	If $\Lambda_i\in \mathcal{L}_i(i=1,2,\ldots)$, then  
	$$\mathcal{M}\left\{\prod_{i=1}^{\infty}\Lambda_i\right\}=
	\bigwedge_{i=1}^{\infty}\mathcal{M}\{\Lambda_i\}.$$
    
For each Borel set $B\subseteq\mathbb{R}$,  $\xi$ is called an \textit{uncertain variable} if the set
$\{\gamma \in \Gamma \mid \xi(\gamma) \in B\} \in \mathcal{L}$.  For any $z\in\mathbb{R}$, the function
$\Phi(z;\boldsymbol{\theta}) = \mathcal{M}\{\xi \le z\}$
is called the \textit{uncertainty distribution} of $\xi$ with unknown parameter vector $\boldsymbol{\theta} (\in \Theta)$.   If $\Phi(x)$ is a continuous, strictly increasing function satisfying
	$\lim_{x \to -\infty} \Phi(x) = 0$, $ \lim_{x \to +\infty} \Phi(x) = 1,$
	then $\Phi(x)$ is called a \emph{regular uncertainty distribution}. Under regularity conditions, the inverse function of $\Phi(z;\boldsymbol{\theta})$ is called the \textit{inverse uncertainty distribution}, denoted by $\Phi^{-1}(\alpha)$ for $\alpha\in(0,1)$. The expectation and the variance  of $\xi$ is defined by
	\[
	E(\xi)=\int_0^1\Phi^{-1}(\alpha)\,\mathrm{d}\alpha, \quad V(\xi)=\int_{0}^{1}\left(\Phi^{-1}(\alpha)-E(\xi)\right)^{2} d \alpha
	\]
if the expected value $E(\xi)$ exists.

An uncertain variable $\xi$ is  normal if its uncertainty distribution and inverse function are defined as:
 \vspace{0cm}
 \begin{align*}
 \Phi(z;e,\sigma) &= \left(1 + \exp\left(\frac{\pi(e-z)}{\sqrt{3}\sigma}\right)\right)^{-1}, \quad z \in \mathbb{R},\\
 \Phi^{-1}(\alpha;e,\sigma) &= e+ \frac{\sqrt{3}\sigma}{\pi}\ln\left(\frac{\alpha}{1-\alpha}\right), \quad \alpha \in (0,1),
 \end{align*}
 where $e$ and $\sigma$ are real numbers with $\sigma > 0$, denoted by  $\xi \sim \mathscr{N}(e, \sigma)$. When $e=0$ and $\sigma=1$, $\xi$ is a standard normal uncertainty distribution and $\xi\sim \mathscr{N}(0, 1)$,	
where $\mu$ and $\sigma(>0)$ are real numbers.  The uncertain normal distribution $\mathcal{N}(e,\sigma)$ is a regular uncertainty distribution.

\begin{lemma}\label{lem1}
 (\cite{18})  {\rm
Let $\xi_i\ (i=1,2)$ be independent and  $\xi_i\sim\mathcal{N}(e_i,\sigma_i)$. Then, $\xi_1+\xi_2\sim\mathcal{N}(e_1+e_2,\sigma_1+\sigma_2)$. 
Moreover, if $\xi\sim\mathcal{N}(e,\sigma)$, then  $k\xi \sim\mathcal{N}(ke,k\sigma)$ for $k>0$.   }
\end{lemma}

\begin{lemma}\label{lem2}
 (\cite{18}) {\rm
Let $\xi_i\ (i=1,\ldots,n)$ be independent uncertain variables with uncertainty distributions $\Phi_i(x_i;\boldsymbol{\theta}_i)$.
Then, for any real numbers $x_1,x_2,\ldots,x_n$, the uncertain vector $(\xi_1,\xi_2,\ldots,\xi_n)$ has the joint uncertainty distribution
$\Phi(x_1,x_2,\ldots,x_n)=\bigwedge_{i=1}^{n}\Phi_i(x_i;\boldsymbol{\theta}_i).
$}
\end{lemma}

\begin{lemma}\label{lem3}
 (\cite{18}) {\rm
Assume $\boldsymbol{\zeta}$ is a normal uncertain vector defined above, $c$ is a real vector, and $D$ is a real matrix.  Then, $\boldsymbol{\eta} = c + D\boldsymbol{\zeta}$ is also a normal uncertain vector.}
\end{lemma}

%%%%%%%%%%%%%%%%%%%%%%%%%%%%%%%%%%%%%%%%%%%%%%
\subsection{ Uncertain Fixed-effects Model}
%%%%%%%%%%%%%%%%%%%%%%%%%%%%%%%%%%%%%%%%%%%%%%

A $k \times k$ Latin square design is formed by a treatment factor $T$ and two block factors $A$ and  $B$,  each with $k$ levels.  The block factors $A$ and $B$ are often called row and column blocks, respectively, and their levels are denoted by $i,j=1,\ldots,k$. The level $l(l=1,\ldots,k)$ of the treatment $T$ appears only once in each row and each column, thus is uniquely determined by the  $i$th  row and the $j$th column, denoted by $l=l(i,j)$. Let $y_{lij}$ be the observations to the $l$th level of treatment  in the $i$th $(i=1,\ldots,k)$  row and the $j$th $(j=1,\ldots,k)$ column of the Latin square arrangement. Let $\tau_l$ be the $l$th effect of the treatment $T$,  and  $a_i$ and $b_j$ be the $i$th, $j$th effect of the row  $A$, and column $B$, respectively. To analyze the observations based on uncertain statistics, we need the following assumptions.

\textit{Assumption 1}. Under treatment and block factors, the observations $\{y_{lij}\}(l=l(i,j), i,j=1,\ldots,k)$ are independnet and follows a  normal uncertainty distribution $ \mathscr{N}(0, \sigma)$. Normally distributed uncertainty remains closed under linear operations, offering mathematical tractability and simplifying the complexity of uncertainty statistical inference. 

\textit{Assumption 2}. The effects  $\tau_l$, $a_i$, and $b_j$ $(l=l(i,j), i,j=1,\ldots,k)$ are fixed and independent of each other.  The levels of both treatment factors and blocking factors are actively set by the researcher, with the research objective focused on comparing the specific effects of the observed treatment levels.

\textit{Assumption 3}. The interaction effects between treatment and block factors are ignorable. This makes the existence of interaction effects lack a physical basis, and there are no systematic associations among observations or among effects. The structure and allocation scheme of the Latin square design ensure the physical separation of experimental units, thereby eliminating systematic associations between treatments and blocks. 

Under Assumptions 1-3, the observations $\{y_{lij}\}$ are described by a linear additive model
\begin{eqnarray}\label{EFmodel}
\left\{\begin{array}{ll}
	\displaystyle y_{lij} = \mu + \tau_l + a_i + b_j + \varepsilon_{ijl}, l=l(i,j), i,j=1,\ldots,k,\\
	\displaystyle \sum_{l=1}^{k} \tau_l=\sum_{i=1}^{k} a_i =\sum_{j=1}^{k} b_j= 0,\\
\displaystyle\varepsilon_{lij} \sim \mathcal{N}(0, \sigma),l=l(i,j), i,j=1,\ldots,k,
\end{array}\right.
\end{eqnarray}
where $\varepsilon_{lij}((l=l(i,j), i,j=1,\ldots,k))$ are independent and identically distributed. 
	The model (\ref{EFmodel}) is called  \textit{ an uncertain fixed-effects (UFE) model} under the Latin square design, where the error term is modeled as an uncertain variable, i.e., non-stochastic uncertainty characterized by an expert’s confidence function.
\begin{remark}
Within the framework of uncertainty theory, observations and disturbance terms are regarded as uncertain variables. The disturbance term (error term) is often assumed to have zero expectation, that is, $E(\varepsilon_{lij})=0$. This assumption is not merely a natural mathematical consequence but a crucial prerequisite for model identification and statistical inference.
If the expectation of the disturbance term is not zero, for example, $E(\varepsilon_{lij} )=\mu' \neq 0$, then	the overall mean $\mu$ would be confounded with the expectation of the disturbance term in the model (\ref{EFmodel}). In this situation, $\mu$ would be absorbed into the error term, making it impossible to distinguish treatment effects $\tau_l$ from the disturbance component, which would in turn lead to systematic bias in parameter estimation.  
\end{remark}

%%%%%%%%%%%%%%%%%%%%%%%%%%%%%%%%%%%%%%%%
\section{Parameter Estimation}\label{s3}
In this section, three methods are proposed to estimate the effects of treatment and block factors in the model (\ref{EFmodel}), including least squares, maximum likelihood, and least absolute deviation estimations. Based on the estimators, we construct the confidence intervals of all effects. In the model (\ref{EFmodel}), there are a total of $3k+1$ unknown parameters. Denote $\bm{\beta} = (\mu, \tau_1, \ldots, \tau_k, a_1, \ldots, a_k, b_1, \ldots, b_k)^{\mathrm{T}}$. 
%%%%%%%%%%%%%%%%%%%%%%%%%%%%%%%%%%%%%%%%
%\subsection{Least Squares Estimation}
%%%%%%%%%%%%%%%%%%%%%%%%%%%%%%%%%%%%%%%%

\textit{Least Squares Estimation (LSE).} Let $\hat{\bm{\beta}}$ and $\hat\sigma$ be the least squares estimators  of  $\bm{\beta}$ and $\sigma$. Since  the observed value $y_{lij}$ is an uncertain response variable, the LSE  $\hat{\bm{\beta}}$ of the parameter $\bm{\beta}$ is the solution to the following optimization problem:
\begin{align}\label{lse}
\hat{\bm{\beta}}=\argmin_{\beta}  
\sum_{i=1}^k\sum_{j=1}^k E[(y_{lij} - (\mu + \tau_{l}+ a_i + b_j ))^2].
\end{align}
The equation (\ref{lse}) is equivalent to minimizing the equation below:
$$
\hat{\bm{\beta}}=\argmin_{\beta}\sum_{i=1}^k\sum_{j=1}^k\left(E\left[y_{lij}\right] - \left(\mu + \tau_l + a_i + b_j\right)\right)^2 + \sum_{i=1}^k\sum_{j=1}^k V\left[y_{lij}\right].
$$

Since the variance term \(V\left[y_{lij}\right]\) does not depend on the parameters, the optimization problem simplifies to:
\[
\hat{\bm{\beta}}=\argmin_{\beta} \sum_{i,j} \left(E\left[y_{lij}\right] - \mu - \tau_l - a_i - b_j\right)^2.
\]
Owing to the linearity of the uncertain expectation, the closed-form estimators $\hat{\bm{\beta}}$ of $\bm{\beta}$ are given by:
\[
\begin{aligned}
	\hat{\mu}     &= \frac{1}{k^2}\sum_{i=1}^{k}\sum_{j=1}^{k} E\left[y_{lij}\right],  \qquad \hat{\tau}_l   = \frac{1}{k}\sum_{l=1}^{k} E\left[y_{lij}\right] - \frac{1}{k^2}\sum_{i=1}^{k}\sum_{j=1}^{k} E\left[y_{lij}\right],\\
	\hat{a}_i      &= \frac{1}{k}\sum_{j=1}^{k} E\left[y_{lij}\right] - \frac{1}{k^2}\sum_{i=1}^{k}\sum_{j=1}^{k} E\left[y_{lij}\right], 
	\quad\hat{b}_j     = \frac{1}{k}\sum_{i=1}^{k} E\left[y_{lij}\right] - \frac{1}{k^2}\sum_{i=1}^{k}\sum_{j=1}^{k} E\left[y_{lij}\right].
\end{aligned}
\]

Following the equation (7) of Lio and Liu (2018b), uncertain residuals of the $(l,i,j)$th observation are defined as:
\begin{equation*}
	\hat{\varepsilon}_{lij} = y_{lij} -(\hat{\mu} + \hat{\tau}_{l} + \hat{a}_i + \hat{b}_j), l=l(i,j), i,j=1,\ldots,k,
\end{equation*}
where $\hat{\mu}, \hat{\tau}_l, \hat{a}_i, \hat{b}_j$ are the parameter estimates obtained via uncertain least squares. Since the model assumes $E[\varepsilon_{lij}]=0$, following the variance estimation framework of Lio and Liu (2018b), the estimate of the disturbance variance ${\hat{\sigma}^2}$ is:
\begin{equation*}
	\hat{\sigma}^2 = \frac{1}{(k-1)(k-2)}\sum_{i=1}^{k}\sum_{j=1}^{k}E[\hat{\varepsilon}_{lij}^{2}],
\end{equation*}
where the denominator $(k-1)(k-2)=k^2-3k+2$ is the correct degrees of freedom for the Latin square design, accounting for the $3k-2$ estimated parameters (including constraints).	
Define the set
$$
\Omega_l = \{(i,j): \text{treatment } l \text{ is assigned to the } (i,j)\text{th experimental unit}\}.
$$
The standard deviation estimators for each substructure are given by
\begin{align*}
	\hat{\sigma}_{a_i}^2 &= \frac{1}{k}\sum_{j=1}^{k}\hat{\varepsilon}_{lij}^{2},\qquad
	\hat{\sigma}_{b_j}^2 = \frac{1}{k}\sum_{i=1}^{k}\hat{\varepsilon}_{lij}^{2},\qquad
	\hat{\sigma}_{\tau_l}^{2} = \frac{1}{k}\sum_{(i,j)\in\Omega_l}\hat{\varepsilon}_{lij}^{2}
\end{align*}
%%%%%%%%%%%%%%%%%%%%%%%%%%%%%%%%%%%%%%%%%%%%%%%%%
%\subsection{ Maximum Likelihood Estimation}
%%%%%%%%%%%%%%%%%%%%%%%%%%%%%%%%%%%%%%%%%%%%%%%%%

\textit{Maximum Likelihood Estimation (MLE).} Let $\tilde{\bm{\beta}}$ and $\tilde \sigma$ be the maximum likelihood estimators  of $\bm{\beta}$ and $\sigma$. Define the residuals:
	\begin{equation*}
		\varepsilon_{lij}(\boldsymbol{\beta}) = y_{lij} - (\mu + \tau_{l} + a_i + b_j), l=l(i,j), i,j=1,\ldots,k,.
	\end{equation*}
  The MLE  $\tilde{\bm{\beta}}$ is  the solution to the following Chebyshev regression problem:
	\begin{align*}
		&\boldsymbol{\tilde{\beta}} = \argmin_{\bm{\beta}} \bigvee_{i,j=1}^{k} \bigl|\varepsilon_{ij}(\boldsymbol{\beta})\bigr|
		= \arg\min_{\bm{\beta}} \bigvee_{i,j=1}^{k} \bigl|y_{lij} - (\mu + \tau_{l} + a_i + b_j)\bigr|,
	\\
	&	\text{s.t. } \sum_{l=1}^{k}\tau_l = \sum_{i=1}^{k}a_i = \sum_{j=1}^{k}b_j = 0.
	\end{align*}
	Transform the Chebyshev regression problem into a standard linear programming form:
	\begin{align*}
		&\tilde{\bm{\beta}}=\argmin_{\bm{\beta}}  R({\bm{\beta}})=\argmin_{\bm{\beta}} \bigvee_{i,j=1}^{k} \bigl|y_{lij} - ({\mu} + {\tau}_{l}+{a_i} + {b_j})\bigr|,
	\\
		&\text{s.t. } 
		\begin{cases}
			\bigl|y_{lij} - (\mu + \tau_{l} + a_i + b_j)\bigr| \leq  R({\bm{\beta}}), \quad i,j=1,\ldots,k, \\[8pt]
			\displaystyle\sum_{l=1}^{k}\tau_l = 0,\quad \sum_{i=1}^{k}a_i = 0,\quad \sum_{j=1}^{k}b_j = 0.
		\end{cases}
	\end{align*}
This linear program can be directly solved using off-the-shelf solvers, e.g.,  \texttt{scipy.optimize.linprog}, \texttt{HiGHS},  \texttt{Gurobi}, or \texttt{CPLEX}, to obtain the global optimal solution $\boldsymbol{\tilde{\beta}}$ and the optimal value $R(\boldsymbol{\tilde{\beta}})$.
	
	Denote the maximum absolute residual as:
	\begin{equation*}
		R(\boldsymbol{\tilde{\beta}}) = \bigvee_{i,j=1}^{k} \bigl|y_{lij} - (\tilde{\mu} + \tilde{\tau}_{l}+\tilde{a_i} + \tilde{b}_j)\bigr|.
	\end{equation*}
	Substituting $R(\boldsymbol{\tilde{\beta}})$ into the likelihood function, $\tilde{\sigma}$ is the solution to the following maximization problem:
	\begin{equation*}
		\tilde{\sigma} = \arg\max_{\sigma>0} \frac{\dfrac{\pi}{\sqrt{3}\sigma}\exp\!\left(\dfrac{\pi}{\sqrt{3}\sigma}R(\boldsymbol{\tilde{\beta}})\right)}{\left[1+\exp\!\left(\dfrac{\pi}{\sqrt{3}\sigma}R(\boldsymbol{\tilde{\beta}})\right)\right]^2}.
	\end{equation*}
For convenience, let $t = \pi R(\boldsymbol{\tilde{\beta}})/(\sqrt{3}\sigma)$. Since $\sigma > 0$ and $R(\boldsymbol{\tilde{\beta}}) > 0$, we have $t > 0$, and $\tilde\sigma = \pi R(\boldsymbol{\tilde{\beta}})/(\sqrt{3}t)$. The original problem is transformed into an optimization problem in $t$. The objective function becomes $g(t) = t e^t(1+e^t)^{-2}$. Differentiating $\ln g(t) = \ln t + t - 2\ln(1+e^t)$ and setting the derivative to zero yields the equation
	\begin{equation*}
		\frac{1}{t^*} + 1 = \frac{2}{1+e^{-t^*}}.
	\end{equation*}
	Solving numerically using Newton's method gives $t^* \approx 1.5434$. Substituting back via $\tilde\sigma = \pi R(\boldsymbol{\tilde{\beta}})/(\sqrt{3}t)$ yields
	\begin{equation*}
		\tilde{\sigma} = \frac{\pi R(\boldsymbol{\tilde{\beta}})}{\sqrt{3}\, t^*} \approx 1.178 R(\boldsymbol{\tilde{\beta}}).
	\end{equation*}
The standard deviation estimators for each substructure are defined as
\begin{align*}
	\tilde{\sigma}_{a_i} &= \frac{\pi}{\sqrt{3}\,t^*}\max_{1\le j\le k}\left|\tilde{\varepsilon}_{lij}\right|,\quad
	\tilde{\sigma}_{b_j} = \frac{\pi}{\sqrt{3}\,t^*}\max_{1\le i\le k}\left|\tilde{\varepsilon}_{lij}\right|,\quad
	\tilde{\sigma}_{\tau_l} = \frac{\pi}{\sqrt{3}\,t^*}\max_{1\le l\le k}\left|\tilde{\varepsilon}_{lij}\right|.
\end{align*}

%%%%%%%%%%%%%%%%%%%%%%%%%%%%%%%%%%%%%%%%%%%%%%%%%
%\subsection{Least Absolute Deviation Estimation}
%%%%%%%%%%%%%%%%%%%%%%%%%%%%%%%%%%%%%%%%%%%%%%%%%
\textit{Least Absolute Deviation (LAD) Estimation.}  Let $\breve{\bm{\beta}}$ and $\breve \sigma$ be the least absolute deviation estimators  of $\bm{\beta}$ and $\sigma$. In the model (\ref{EFmodel}),  
${y}_{lij} = (\mu+ \tau_l+ a_i+ b_j) + \varepsilon_{lij}$ and $ \varepsilon_{lij} \sim \mathcal{N}(0,\breve\sigma)
$.
Thus, by Lemma \ref{lem1}, the uncertainty distribution of the observed variable is
\begin{equation*}
{y}_{lij} \sim \mathcal{N}(\theta_{lij}, \sigma), \quad \theta_{lij}= \mu+ \tau_l+ a_i+ b_j, l=l(i,j), i,j=1,\ldots,k,.
\end{equation*}
Its inverse uncertainty distribution is of the following type:
\begin{equation}\label{iud}
	\Psi_{lij}^{-1}(\alpha) = \theta_{lij}+ \frac{\sigma\sqrt{3}}{\pi}\ln\frac{\alpha}{1-\alpha}.
\end{equation}

Following the uncertain least absolute deviation estimation framework of Lio and Liu \cite{20}, the estimator  $ {\bm{\breve\beta}}$ can be  performed by minimizing the expected absolute deviation:
\begin{equation*}
	{\bm{\breve\beta}}=\argmin_{{\bm{\beta}}} \sum_{i=1}^k\sum_{j=1}^k \mathbb{E}\left|y_{lij} - (\mu + \tau_{l} + a_i + b_j)\right|.
\end{equation*}
Using the integral representation of the expectation of an uncertain variable, the optimization problem is equivalent to:
\begin{equation}\label{eq:objective}
	{\bm{\breve\beta}}=\argmin_{{\bm{\beta}}} \sum_{i=1}^k\sum_{j=1}^k \int_0^1 \left|\Psi_{lij}^{-1}(\alpha) - (\mu + \tau_{l} + a_i + b_j)\right| d\alpha
\end{equation}
subject to the constraints:
\begin{equation*}\label{eq:constraints}
	\sum_{l=1}^k \tau_l = 0, \quad \sum_{i=1}^k a_i = 0, \quad \sum_{j=1}^k b_j = 0,
\end{equation*}
where $\Psi_{lij}^{-1}(\alpha)$ is defined in (\ref{iud}).

In this study, we adopt a linear programming (LP) transformation to solve the problem. The objective function involves a non‑smooth absolute value and an integral, while the decision variables are subject to linear equality constraints; this constitutes a constrained non‑smooth convex optimization problem. The LP transformation reformulates the original problem into a standard linear program via equivalent transformations, yielding the global optimum and enabling efficient solution with mature optimization software. It is the optimal choice that balances theoretical rigor and computational feasibility.

We apply the composite trapezoidal rule in open‑interval form to avoid endpoint singularities and discretize $[0,1]$ into $M$ nodes $\alpha_m=m/(M+1)$ ($m=1,\ldots, M$) with weights:
\begin{equation*}
	w_m=
	\begin{cases}
		(2(M+1))^{-1}, & m=1,M, \\[6pt]
		{(M+1)^{-1}}, & m=2,\ldots,M-1.
	\end{cases}
\end{equation*}
Define the discretized inverse distribution values determined by the true parameters $\theta_{lij}$:
\begin{equation*}
	\Psi_{ijm}:=\Psi_{lij}^{-1}(\alpha_m)=\theta_{lij}+\eta\ln\frac{\alpha_m}{1-\alpha_m}, 
\end{equation*}
where $\eta=\frac{\sigma\sqrt{3}}{\pi}$. The original objective function (\ref{eq:objective}) is approximated by:
\begin{equation}\label{eq:discrete}
{\bm{\breve\beta}}=	\argmin_{{\bm{\beta}}}\sum_{i=1}^k\sum_{j=1}^k\sum_{m=1}^Mw_m\left|\Psi_{ijm}-(\mu+\tau_{l}+a_i+b_j)\right|.
\end{equation}
For each triple $(i,j,m)$,  auxiliary variables $u_{ijm}\geq0$, $v_{ijm}\geq0$ are introduced such that:
\begin{equation*}\label{eq:decomposition}
	\Psi_{ijm}-(\mu+\tau_{l}+a_i+b_j)=u_{ijm}-v_{ijm}.
\end{equation*}
By the complementary slackness theory of linear programming, at the optimum we have $u_{ijm}\cdot v_{ijm}=0$, and then $|u_{ijm}-v_{ijm}|=u_{ijm}+v_{ijm}$. The objective function (\ref{eq:discrete}) is equivalent to the linear form:
\begin{equation}\label{eq:linear}
{\bm{\breve\beta}}=	\argmin_{{\bm{\beta}}}\sum_{i=1}^k\sum_{j=1}^k\sum_{m=1}^Mw_m(u_{ijm}+v_{ijm}).
\end{equation}
We eliminate redundant parameters via the equality constraints to strictly satisfy the identifiability conditions. Denote
\begin{equation}\label{eq:reduction}
	\tau_k=-\sum_{l=1}^{k-1}\tau_l,\quad a_k=-\sum_{i=1}^{k-1}a_i,\quad b_k=-\sum_{j=1}^{k-1}b_j.
\end{equation}
Define the reduced parameters according to three cases:
(i) if $l\leq k-1$, keep $\tau_{l}$; if $l(i,j)=k$, replace with $-\sum_{p=1}^{k-1}\tau_p$,  
(ii) if $i\leq k-1$, keep $a_i$; if $i=k$, replace with $-\sum_{p=1}^{k-1}a_p$, and 
(iii) if $j\leq k-1$, keep $b_j$; if $j=k$, replace with $-\sum_{p=1}^{k-1}b_p$. After this transformation, we obtain the standard LP of (\ref{eq:linear}):
\begin{equation}\label{eq:lp}
	\begin{aligned}
		&{\bm{\breve\beta}}=	\argmin_{{\bm{\beta}}}  \sum_{i=1}^k\sum_{j=1}^k\sum_{m=1}^M w_m u_{ijm} + \sum_{i=1}^k\sum_{j=1}^k\sum_{m=1}^M w_m v_{ijm}， \\
		&\mathrm{s.t.} \left\{\begin{array}{ll}
   \mu+\tau_{l_{ij}}^{\dagger}+a_i^{\dagger}+b_j^{\dagger}+u_{ijm}-v_{ijm}=\Psi_{ijm},& \forall\,i,j,m, \\[8pt]
		  u_{ijm}\geq0,\ v_{ijm}\geq0, & \forall\,i,j,m,
\end{array}\right.
	\end{aligned}
\end{equation}
where $\tau_{l_{ij}}^{\dagger},a_i^{\dagger},b_j^{\dagger}$ are the reduced expressions, and $\mu,\tau_l,a_i,b_j$, with $l,i,j\leq k-1$, are free variables.

We call a standard LP solver (e.g., HiGHS, Gurobi) to solve the linear programming problem formulated above. After obtaining the global optimal solution, we recover the original parameters as follows: 
(i) $\breve{\mu}$ is  directly read,
(ii) $\breve{\tau}_l$ ($l=1,\ldots,k-1$) are directly read, and $\breve{\tau}_k=-\sum_{l=1}^{k-1}\breve{\tau}_l$,
(iii) $\breve{a}_i$ ($i=1,\ldots,k-1$) are directly read, and $\breve{a}_k=-\sum_{i=1}^{k-1}\breve{a}_i$, and
(iv) $\breve{b}_j$ ($j=1,\ldots,k-1$) are directly read, and $\breve{b}_k=-\sum_{j=1}^{k-1}\breve{b}_j$.

Let the location parameters estimated by LP be $\breve{\theta}_{lij}=\breve{\mu}+\breve{\tau}_{l(i,j)}+\breve{a}_i+\breve{b}_j$. The estimated residual is
$$\breve{\varepsilon}_{ij}={y}_{lij}-\breve{\theta}_{lij}.$$
Its inverse uncertainty distribution is $\Psi_{\breve{\varepsilon}_{lij}}^{-1}(\alpha)=\Psi_{lij}^{-1}(\alpha)-\breve{\theta}_{lij},$
where $\Psi_{lij}^{-1}(\alpha)$ is defined in (\ref{iud}).
Since the variance of an uncertain variable can be precisely expressed via the second-order moment integral of its inverse distribution function, the variance estimator is given by
\begin{equation*}
	\breve{\sigma}^2=\frac{1}{(k-1)(k-2)}\sum_{i=1}^k\sum_{j=1}^k\int_0^1\left(\Psi_{lij}^{-1}(\alpha)-\breve{\theta}_{ij}\right)^2d\alpha.
\end{equation*}                
Discretizing the computation yields
\begin{equation*}
	\breve{\sigma}^2=\frac{1}{(k-1)(k-2)}\sum_{i,j}\sum_{m=1}^Mw_m\left(\Psi_{ijm}-\breve{\theta}_{lij}\right)^2,
\end{equation*}
where $\Psi_{ijm}=\Psi_{lij}^{-1}(\alpha_m)$, and the denominator $(k-1)(k-2)$ is the error degrees of freedom for the Latin square design.
The variance estimators for each substructure are given by
\begin{align*}
	\breve{\sigma}_{a_i}^2 
	&= \frac{1}{k}\sum_{j=1}^{k}\sum_{m=1}^{M} w_m \left(\Psi_{ijm}-\breve{\theta}_{lij}\right)^2,
	\quad
	\breve{\sigma}_{b_j}^2 
	= \frac{1}{k}\sum_{i=1}^{k}\sum_{m=1}^{M} w_m \left(\Psi_{ijm}-\breve{\theta}_{lij}\right)^2, \\
	\breve{\sigma}_{\tau_l}^2 
	&= \frac{1}{k}\sum_{l=1}^{k}\sum_{m=1}^{M} w_m \left(\Psi_{ijm}-\breve{\theta}_{lij}\right)^2.
\end{align*}

%%%%%%%%%%%%%%%%%%%%%%%%%%%%%%%%%%%%%%%%%%%%%%%%%
%\subsection{Confidence Intervals}
%%%%%%%%%%%%%%%%%%%%%%%%%%%%%%%%%%%%%%%%%%%%%%%%%
\textit{ Confidence Intervals based on Three Estimation Methods.} Let the predicted uncertain variable be defined as: 
$	 \overset{*}{y}= \overset{*}{\mu} + \overset{*}{\tau}_l + \overset{*}{a}_i + \overset{*}{b}_j + \varepsilon,$
where $\varepsilon \sim \mathcal{N}(0, \overset{*}{\sigma})$, $\overset{*}{\varepsilon} = y - \overset{*}{y} $ , and $\overset{*}{c}=\hat{c}, \tilde{c}, \breve{c}$. 
	The predicted value of the response variable is taken as the expected value of the predicted uncertain variable:
$
	\tilde{\mu} = \overset{*}{\mu} + \overset{*}{\tau}_l + \overset{*}{a}_i + \overset{*}{b}_j.
$	
	Consequently, the predicted uncertain variable $\overset{*}{y}$ follows the normal uncertainty distribution $\mathcal{N}(\tilde{\mu}, \overset{*}{\sigma})$, and the corresponding inverse uncertainty distribution function is given by:
	\[
	\overset{*}{\Psi}^{-1}(\alpha) = \tilde{\mu} + \frac{\overset{*}{\sigma}\sqrt{3}}{\pi}\ln\frac{\alpha}{1-\alpha}, \quad \alpha \in (0,1).
	\]
	For a confidence level $\alpha \in (0,1)$, by the subadditivity of the uncertainty measure, the $(1-\alpha)$ confidence interval for the response variable $y$ can be expressed as:
	\[
	\left[\overset{*}{\Psi}^{-1}\left(\frac{1-\alpha}{2}\right),\ \overset{*}{\Psi}^{-1}\left(\frac{1+\alpha}{2}\right)\right].
	\]
	Substituting the inverse uncertainty distribution function yields the explicit expression:
	\[
	\left[\tilde{\mu} + \frac{\overset{*}{\sigma}\sqrt{3}}{\pi}\ln\frac{1-\alpha}{1+\alpha},\ 
	\tilde{\mu} + \frac{\overset{*}{\sigma}\sqrt{3}}{\pi}\ln\frac{1+\alpha}{1-\alpha}\right].
	\]
	For a $k$-order Latin square design, the general form of the $(1-\alpha)$ confidence intervals for the parameters is:
\begin{align*}
	CI(\mu) &= \overset{*}{\mu} \pm \frac{\overset{*}{\sigma}_{\mu}\sqrt{3}}{\pi}\ln\frac{1+\alpha}{1-\alpha}, \quad
	CI(a_i) = \overset{*}{a}_i \pm \frac{\overset{*}{\sigma}_{a_i}\sqrt{3}}{\pi}\ln\frac{1+\alpha}{1-\alpha}, \\[6pt]
	CI(b_j) &= \overset{*}{b}_j \pm \frac{\overset{*}{\sigma}_{b_j}\sqrt{3}}{\pi}\ln\frac{1+\alpha}{1-\alpha},  \quad
	CI(\tau_l) = \overset{*}{\tau}_l \pm \frac{\overset{*}{\sigma}_{\tau_l}\sqrt{3}}{\pi}\ln\frac{1+\alpha}{1-\alpha},
\end{align*}
	where $\overset{*}{\sigma}_{\mu}, \overset{*}{\sigma}_{a_i}, \overset{*}{\sigma}_{b_j}, \overset{*}{\sigma}_{\tau_l}$ are the estimated standard deviations of the respective estimators.
%%%%%%%%%%%%%%%%%%%%%%%%%%%%%%%%
\section{ Homogeneity and Common Tests}\label{s4}
%%%%%%%%%%%%%%%%%%%%%%%%%%%%%%%%%%%%
In this section, we conduct homogeneity and common tests of standard deviation, treatment effects, and the fitted model.

\textit{Standard Deviation Testing.} 
In the uncertain model (1), we assume that for any $i,j$, $\varepsilon_{lij} \sim \mathcal{N}(0,\sigma)$. In empirical analysis, it is necessary to test whether the standard deviations corresponding to the observations $y_{lij}$ are homogeneous. The homogeneity of standard deviations is decomposed into three orthogonal dimensions: treatment, row, and column, for diagnostic purposes.

To ensure the test focuses solely on dispersion, the row, column, and treatment effects are removed. The adjusted residual is defined as:
\begin{equation*}
	\overset{*}e_{lij} = y_{lij} - (\overset{*}{\mu} + \overset{*}{a}_i + \overset{*}{b}_j + \overset{*}{\tau}_{l})
\end{equation*}
so that $\overset{*}e_{lij}$ approximately follows a zero-mean distribution. According to the experimental structure, the residuals are projected into three grouped dimensions:
(i) Treatment: $G_l^{(T)} = \{\overset{*}e_{lij} \mid l \text{ is fixed}\}$; (ii) Row: $G_i^{(R)} = \{\overset{*}e_{lij} \mid i \text{ is fixed}\}$; and   
Column: $G_j^{(C)} = \{\overset{*}e_{lij} \mid j \text{ is fixed}\}$.
For each dimension $d \in \{T,R,C\}$, the homogeneity test of standard deviations is formulated as:
\begin{equation}
	H_0^{(d)}: \sigma_1^{(d)} = \cdots = \sigma_k^{(d)} \quad \text{vs} \quad H_1^{(d)}: \sigma_1^{(d)}  \text{not all equal}.
\end{equation}
Let $\gamma = 1 - \alpha$ be the coverage probability, and define the quantile coefficient
$	K_\gamma = \frac{\sqrt{3}}{\pi}\ln\left(\frac{2-\alpha}{\alpha}\right)
$,
$\overset{*}{\sigma}_r^{(d)}$ denotes the standard deviation of the $r$th group under dimension $d$. For example, under the row-block dimension, using the uncertain least squares method, the component standard deviation for the $r$-th level is defined as
\[
\hat{\sigma}_{r}^{R} = \sqrt{\frac{1}{k}\sum_{j=1}^{k}\hat{\varepsilon}_{lrj}^2}.
\]
For any two distinct group indices $r, s \in \{1,\ldots,k\}$ ($r \neq s$) under dimension $d$, define the abnormal deviation proportion of group $s$ relative to group $r$, denoted by $
\Pi_{s|r}^{(d)}$, as
\begin{equation}\label{T1}
\Pi_{s|r}^{(d)} = \frac{1}{k}\#\left\{m \in \{1,\ldots,k\} :
e_m^{(s)} < -\overset{*}{\sigma}r^{(d)} K\gamma
\ \text{ or } e_m^{(s)} > \overset{*}{\sigma}r^{(d)} K\gamma
\right\},
\end{equation}
where $\#\{\cdot\}$ denotes the cardinality of a set, and $e_m^{(s)}$ represents the $m$th residual observation in the $s$th group under dimension $d$.
The rejection region for dimension $d$ is defined as
\[
W^{(d)} = \left\{
\mathbf{e} :\exists\ r,s \in \{1,\ldots,k\},\ r \neq s,\ \text{ such that } \Pi_{s|r}^{(d)} \geq \alpha
\right\}.
\]
If the observed residuals fall into $W^{(d)}$, then $H_0^{(d)}$ is rejected, indicating heteroscedasticity in dimension $d$; otherwise, $H_0^{(d)}$ is accepted.

If the multi-dimensional homogeneity tests jointly accept the hypothesis $H_0^{(T)} \cap H_0^{(R)} \cap H_0^{(C)}$, then the disturbance terms in the Latin square model can be assumed to share a common scale parameter, i.e., $\varepsilon_{lij} \sim \mathcal{N}(0, \overset{*}\sigma)$. We further test whether the common standard deviation equals a specified constant $\overset{*}\sigma_0$, formulated as:
\begin{equation}
	H_0^{\sigma}: \overset{*}\sigma = \overset{*}\sigma_0 \quad \text{vs} \quad H_1^{\sigma}: \overset{*}\sigma \neq \overset{*}\sigma_0.
\end{equation}
Let all residuals be pooled into a sequence $\mathbf{E} = \{e_1, \ldots, e_N\}$ with $N = k^2$. Define the number of abnormal observations $M$:
\begin{equation}\label{T2}
	M = \#\left\{ m \in \{1,\ldots,N\} : e_m < -\overset{*}\sigma_0 K_\gamma \ \text{or}\ e_m > \overset{*}\sigma_0 K_\gamma \right\}.
\end{equation}
The rejection region for $H_0^{\sigma}$ is:
$
	W^{\sigma} = \left\{ \mathbf{E} : \frac{M}{N} \geq \alpha \right\}.
$
If the observed residuals fall into $W^{\sigma}$, then $H_0^{\sigma}$ is rejected, indicating that the common standard deviation is not equal to $\sigma_0$; otherwise, $H_0^{\sigma}$ is accepted, and $\overset{*}\sigma = \overset{*}\sigma_0$ is considered valid.
%%%%%%%%%%%%%%%%%%%%%%

\textit{Treatment Effect Testing.} 
Under both homoscedastic and heteroscedastic cases, we study the homogeneity test and the global significance test of treatment effects. To extract the adjusted observations for treatment effect inference, we remove the overall mean, row effect, and column effect from the original observations:
$
	\overset{*}{y}_{lij} = y_{lij} - (\overset{*}{\mu} + \overset{*}{a}_{i} + \overset{*}{b}_{j}).
$
Theoretically, we have $\overset{*}{y}_{lij} \approx \tau_l + \varepsilon_{lij}$. Let the $k$ observations belonging to treatment $l$ be denoted by the set $\overset{*}{Y}^{(l)} = \{\overset{*}{y}_1^{(l)}, \dots, \overset{*}{y}_k^{(l)}\}$, with corresponding parameter estimate $\overset{*}{\tau}_l$.
Under the treatment dimension $T$, based on the least squares method, the component standard deviation for the $r$-th treatment group is given by
\[
\hat{\sigma_{r}}^{T} = \sqrt{\frac{1}{k}  \sum_{(i,j)\in\Omega_l} \hat{\varepsilon}_{rij}^2}.
\]

Next,  we consider the homogeneity hypothesis:
$
H_0^{t}: \tau_1 = \cdots = \tau_k \quad \text{vs} \quad H_1^{t}: \text{not all equal}.
$ There exist the following two cases: 

Case (i): Homoscedasticity using the estimated global standard deviation $\overset{*}{\sigma}$:
\begin{equation}\label{T31}
	\Pi_{s|r}^{(t1)} = \frac{1}{k}\#\left\{m \in \{1,\ldots,k\} :
	\overset{*}y_m^{(l)} < \overset{*}{\tau}_r - \overset{*}{\sigma} K_\gamma
	\ \text{or}\
	\overset{*}y_m^{(l)} > \overset{*}{\tau}_r + \overset{*}{\sigma} K_\gamma
	\right\}.
\end{equation}
The rejection region is $W^{(t1)} = \{ \overset{*}{Y} : \exists\, l \neq r,\ \Pi_{s|r}^{(t1)} \geq \alpha \}$.

Case (ii): Heteroscedasticity using group-specific local standard deviation $\overset{*}{\sigma}_r$:
\begin{equation}\label{T32}
	\Pi_{s|r}^{(t2)} = \frac{1}{k}\#\left\{m \in \{1,\ldots,k\} :
	\overset{*}{y}_m^{(l)} < \overset{*}{\tau}_r - \sigma_{r}^{*(T)} K_\gamma
	\ \text{or}\
	\overset{*}{y}_m^{(l)} > \overset{*}{\tau}_r + \sigma_{r}^{*(T)} K_\gamma
	\right\}.
\end{equation}
The rejection region is $W^{(t2)} = \{ \overset{*}{Y} : \exists\, l \neq r,\ \Pi _{s|r}^{(t2)} \geq \alpha \}$.

If the observed dataset falls into $W^{t}$, then $H_0^{t}$ is rejected, indicating that at least one treatment effect significantly deviates from the common level.

\begin{remark}
	Using the global variance under homoscedasticity yields higher statistical power and more stable confidence intervals, allowing more sensitive detection of small differences among treatment effects $\tau_l$. In the presence of heteroscedasticity, the local-variance-based procedure is more robust and effectively avoids misleading global inference caused by excessive variability within a single group.
\end{remark}
%%%%%%%%%%%%%%5
%\subsubsection{Global Test of Treatment Effects}
%%%%%%%%%%%%%%%%5

Under the homogeneity assumption, we further test whether the overall effect is significantly different from zero:
$
H_0^{tc}: \tau = 0 \quad \text{vs} \quad H_1^{tc}: \tau \neq 0.
$
We denote the pooled observations by $\overset{*}{Y}$, satisfying $\overset{*}{Y} = \overset{*}{Y}^{(l)}$ ($l=1,2,\dots,k$), in other words, $\overset{*}{Y}$ refers to the combined dataset of all samples, with total sample size $N = k^2$. Similarly, two cases are considered: 

Case (i): Homoscedasticity. The rejection region is $W^{(tc1)} = \{\overset{*}{Y} : \Pi^{(tc1)} \geq \alpha \}$, where
\begin{align}\label{T41}
	\Pi^{(tc1)} = \frac{1}{N}\#\{ (l,m) :
	\overset{*}{y}_m^{(l)} < - \overset{*}{\sigma} K_\gamma
	\ \text{or}\
	\overset{*}{y}_m^{(l)} > \overset{*}{\sigma} K_\gamma\}.
\end{align}

Case (ii): Heteroscedasticity. The rejection region is $W^{(tc2)} = \{ \overset{*}{Y} : \Pi^{(tc2)} \geq \alpha \}$, where
\begin{align}\label{T42}
	\Pi^{(tc2)} = \frac{1}{N}\#\{ (l,m) :
	\overset{*}{y}_m^{(l)} < - \overset{*}{\sigma}_l K_\gamma
	\ \text{or}\
	\overset{*}{y}_m^{(l)} > \overset{*}{\sigma}_l K_\gamma\}.
\end{align}
If the observed dataset $\overset{*}{Y}$ falls into $W^{(tc)}$, then $H_0^{tc}$ is rejected, indicating a significant non-zero effect; otherwise, $H_0^{tc}$ is accepted.

%%%%%%%%%%%%%%%%%%%%%%%%%%%%%%%%
\textit{Test of the Fitted Model.}  
Assume that the estimated disturbance term $\varepsilon^*$ follows a normal uncertain distribution $\mathcal{N}(0,\sigma^*)$. Accordingly, the predicted uncertain response variable $y$ is given by
\[
y_{lij}^* = \mu^* + \tau_l^* + a_i^* + b_j^* + \varepsilon^*, 
\quad \varepsilon^* \sim \mathcal{N}(0,\sigma^*).
\]

To assess the adequacy of the distributional assumption for $\varepsilon^*$, we consider the following hypotheses:
\[
H_0^{m}: \sigma = \sigma^* 
\quad \text{vs} \quad 
H_1^{m}: \sigma \neq \sigma^*.
\]
Given a significance level $\alpha$ (set to $0.05$), define the \emph{model deviation proportion} $\Pi^{(m)}$as follows
\[
\Pi^{(m)} = \frac{1}{k^2}
\#
\left\{ (i,j) : \varepsilon_{lij} < \Phi^{-1}\!\left(\frac{\alpha}{2}\right)
\ \text{or}\
\varepsilon_{lij} > \Phi^{-1}\!\left(1 - \frac{\alpha}{2}\right),
\ 1 \le i,j \le k
\right\}.
\]
Based on the estimated residuals
$
\overset{*}\varepsilon_{lij} = y_{lij} - \bigl(\mu^* + \tau_l^* + a_i^* + b_j^*\bigr),
$
following the idea of non-nested distribution family properties, the rejection region is defined as
$
W^{(m)} = \{ (\varepsilon_{lij} : \Pi^{(m)}  \ge \alpha\}.
$
If $\Pi^{(m)} \ge \alpha$, the observed sequence falls into the rejection region $W^{(m)}$, and $H_0^{m}$ is rejected, indicating that either the assumed disturbance distribution $\mathcal{N}(0,\sigma^*)$ or the fitted regression model is inappropriate. Otherwise, $H_0^{m}$ is not rejected, suggesting that both are adequate.
%%%%%%%%%%%%%%%%%%%%%%%%%%%%%%%%%%%
\section{Numerical Simulation and Case Study}\label{s5}
%%%%%%%%%%%%%%%%%%%%%%%%%%%%%%%%%%%
In this section, we first compare the performance of the estimation methods and conduct the testing procedures, and then study a real study.

\subsection{Comparison of Uncertain Estimation Methods}
%%%%%%%%%%%%5

To comprehensively compare the performance of least squares, maximum likelihood estimation, and least absolute deviations estimations within the uncertain Latin square model, this section conducts sensitivity and robustness analyses under varying noise levels, data contamination, and block-effect fluctuations. 

Consider a Latin square design with $k=5$ and $N=3000$ Monte Carlo replications. To ensure model identifiability, both data generation and parameter estimation strictly satisfy the \emph{sum-to-zero constraints}:
$
\sum_{i=1}^{5} a_i = 0,  
\sum_{j=1}^{5} b_j = 0, 
\sum_{l=1}^{5} \tau_l = 0.
$
Take
$\mu = 10$,
$a = -1, -0.5, 0, 0.5, 1$,
$b = -0.4, -0.2, 0, 0.2, 0.4$,
$\tau = -2, -1, 0, 1, 2$.
Under the uncertainty framework, the observations are generated as
\[
y_{lij} = \mu + \tau_l + a_i + b_j + \varepsilon_{ijl}, 
\quad l = l(i,j),\ i,j = 1,\ldots,5,
\]
where $\varepsilon_{ijl}$ denotes the uncertain normal disturbance satisfying $\varepsilon_{ijl} \sim \mathcal{N}(0,\sigma)$. 

To avoid mutual masking of errors across different effects, the evaluation metrics are computed separately for treatment, row, and column effects. For treatment effects $\tau_i$ (or block effects $a_i, b_j$), let the true value be $\theta_i$ and the estimate in the $s$th simulation be $\hat{\theta}_i^{(s)}$. 
The following six core metrics are defined to assess the overall performance of estimation methods across all parameters and simulation runs. Define
 \begin{align*}
\text{Bias} = \frac{1}{N k} \sum_{s=1}^{N} \sum_{i=1}^{k}( \hat{\theta}_i^{(s)} - \theta_i), \quad
\text{MSE} = \frac{1}{Nk} \sum_{s=1}^{N} \sum_{i=1}^{k}( \hat{\theta}_i^{(s)} - \theta_i)^2,\\
\text{MAE} = \frac{1}{Nk} \sum_{s=1}^{N} \sum_{i=1}^{k} |\hat{\theta}_i^{(s)} - \theta_i |, \quad
\text{SD} = \sqrt{\frac{1}{Nk} \sum_{s=1}^{N} \sum_{i=1}^{k}( \hat{\theta}_i^{(s)} - \bar{\hat{\theta}}_i)^2},
\end{align*} 
where $\bar{\hat{\theta}}_i = \frac{1}{N} \sum_{s=1}^{N} \hat{\theta}_i^{(s)}$.
 Bias is used to assess the estimator's unbiasedness.  Mean squared error (MSE) evaluates overall estimation accuracy, with heavier penalties for large deviations. Mean absolute error (MAE) measures the average absolute deviation from the true values and reflects robustness to outliers.  Standard deviation (SD) quantifies the dispersion of estimates around their own mean, reflecting the stability of the estimation method. 
Coverage probability (CP) is the proportion of times the true parameter lies within the confidence interval (target value $0.95$). Average width (AW) is the mean length of the confidence intervals, reflecting estimation precision.

The $(1-\alpha)$ confidence interval for each parameter is given by
$
\text{CI}(\theta) = \overset{*}{\theta} \pm \frac{\overset{*}{\sigma}_{\theta}\sqrt{3}}{\pi}
\ln\frac{1+\alpha}{1-\alpha},
$
where $\overset{*}{\sigma}_{\theta}$ denotes the estimated standard deviation corresponding to the estimator.

\begin{example}\label{exam1}
(Sensitivity Analysis under Varying Noise Levels) 
Assume that the disturbance term $\varepsilon_{lij} \sim \mathcal{N}(0,\sigma)$ with the scale parameter $\sigma \in \{0.1, 0.5, 1.0, 1.5, 2.0, 2.5, 3.0\}$ to investigate the impact of increasing uncertainty (noise level) on estimation accuracy.  Fig. \ref{fig1} shows a comprehensive performance comparison of the three estimation methods under different noise levels.
\begin{figure}[htbp]
	\centering
	\begin{minipage}{0.48\textwidth}
		\centering
		\includegraphics[width=\textwidth]{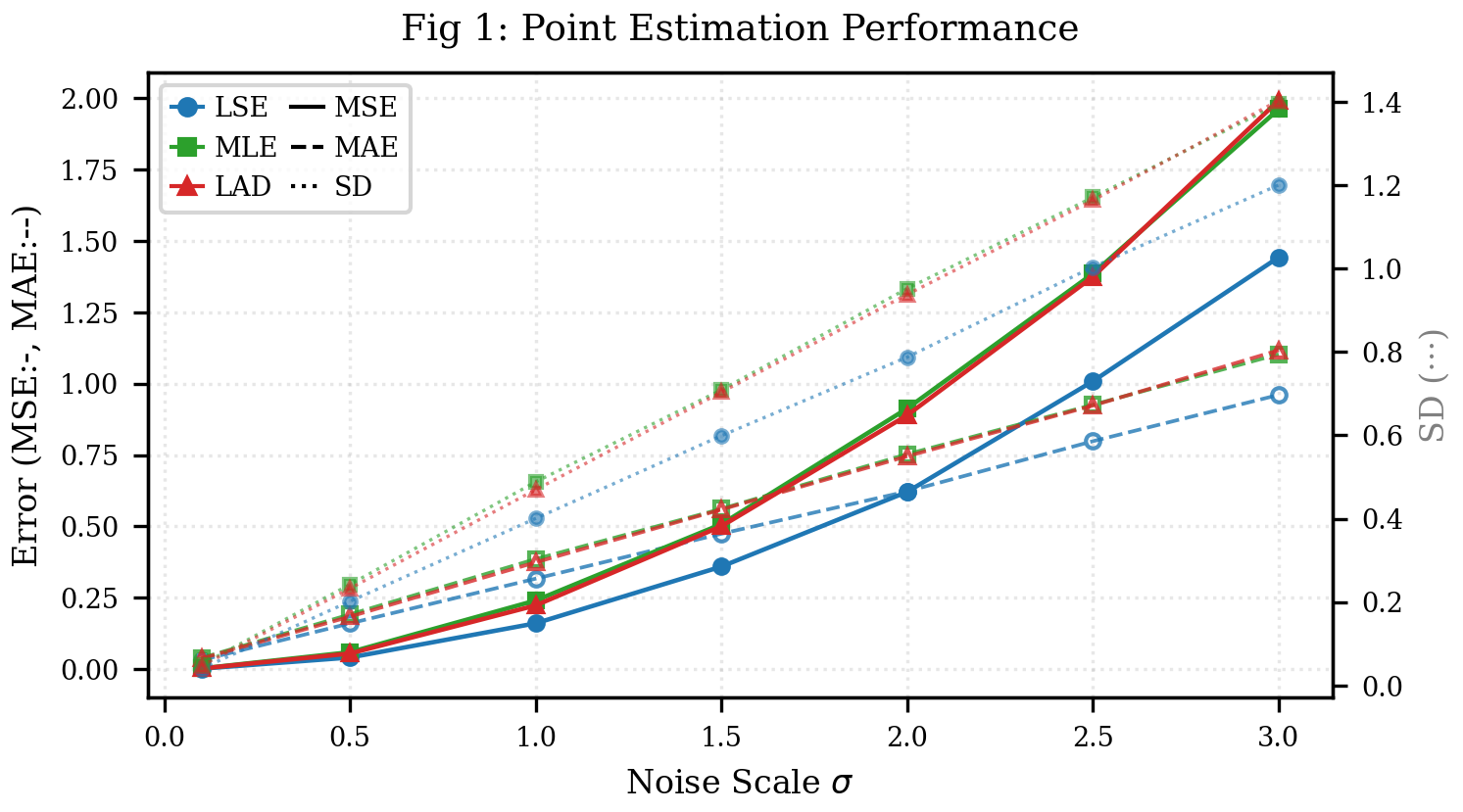}
	\end{minipage}
	\hfill
	\begin{minipage}{0.48\textwidth}
		\centering
		\includegraphics[width=\textwidth]{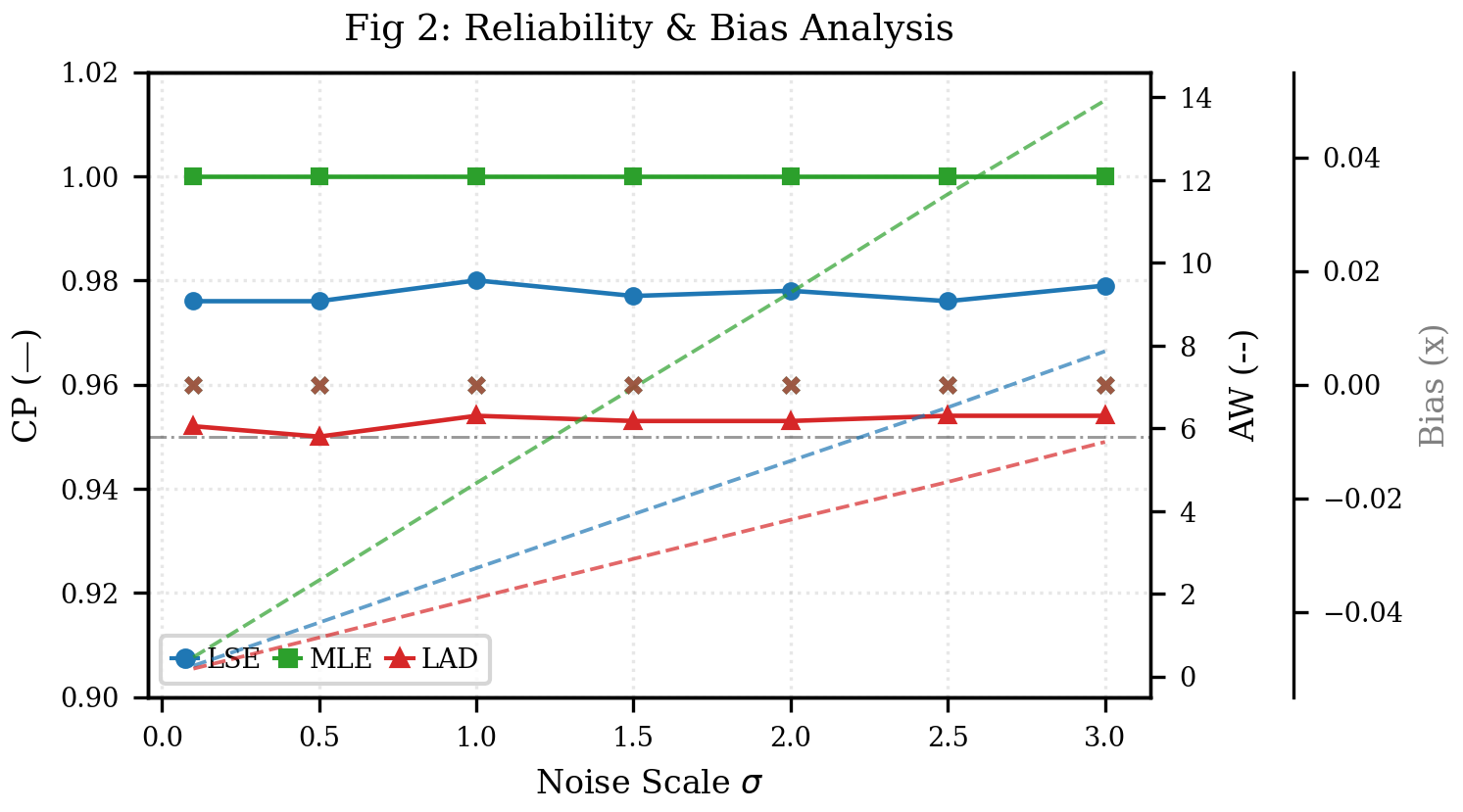}
	\end{minipage}
	\caption{Comprehensive performance comparison of the three estimation methods under varying noise levels.}\label{fig1}
\end{figure}

As the scale parameter $\sigma$ increases, the error metrics of all three estimation methods exhibit a pronounced nonlinear upward trend, while maintaining approximate unbiasedness throughout the entire parameter range. Among them, the MLE demonstrates the weakest robustness under high-noise conditions. Its error measures are consistently larger, and the confidence intervals become substantially inflated in order to maintain the nominal confidence level. For example, when $\sigma = 3.0$, the  $\mathrm{AW}$ reaches $13.93$, indicating an overly conservative behavior that limits its practical applicability. In contrast, the LSE achieves the highest point-estimation accuracy across all noise levels. Both the $\mathrm{MSE}$ and standard deviation remain the lowest among the three methods under low- and high-noise scenarios. In particular, when $\sigma = 3.0$, the $\mathrm{MSE}$ is only $1.44$, demonstrating its strong ability to suppress the adverse effects of increasing variability. The least  LAD, on the other hand, exhibits a distinct advantage in interval estimation. It consistently produces the narrowest confidence intervals while achieving coverage probabilities $\mathrm{CP}$ closest to the nominal level of $0.95$. Specifically, at $\sigma = 3.0$, its average interval width is $5.67$, substantially smaller than that of LSE ($7.86$).Overall, if the primary objective is to obtain highly accurate point estimates, LSE is the preferred choice. However, when the focus is on constructing the most compact and reliable confidence intervals, LAD provides the most favorable performance.
\end{example}
%%%%%%
\begin{example}\label{exam2} (Robustness Analysis under Contaminated Data)  Contaminated data are generated using a mixture uncertain variable model, where the disturbance term follows
$
\varepsilon_{lij} \sim (1-\gamma)\mathcal{N}(0,1) + \gamma \mathcal{N}(0,10)
$
with contamination proportion $\gamma \in \{5\%, 10\%, 15\%, 20\%\}$.

\begin{figure}[htbp]
	\centering
	\begin{minipage}{0.48\textwidth}
		\centering
		\includegraphics[width=\textwidth]{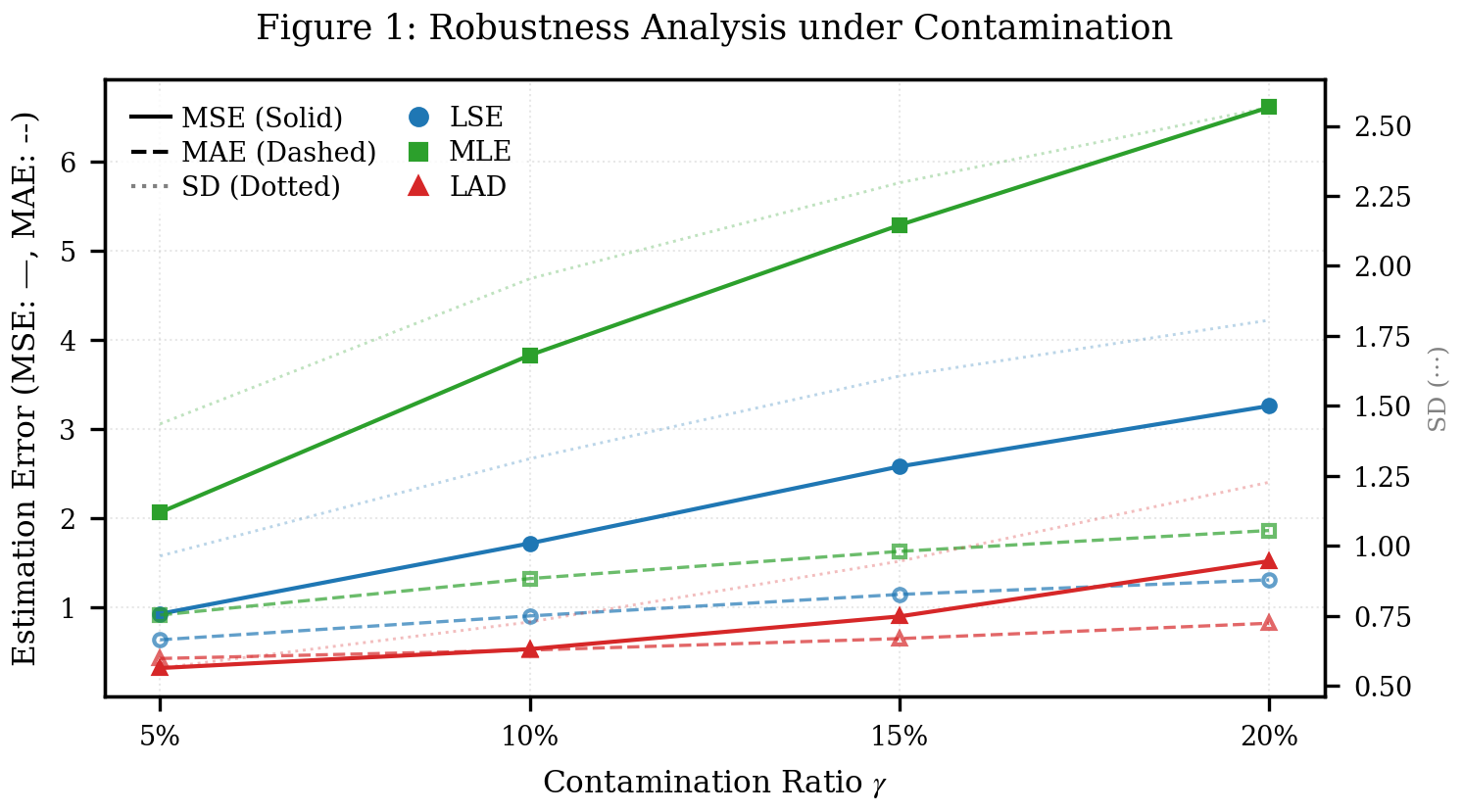}
	\end{minipage}
	\hfill
	\begin{minipage}{0.48\textwidth}
		\centering
		\includegraphics[width=\textwidth]{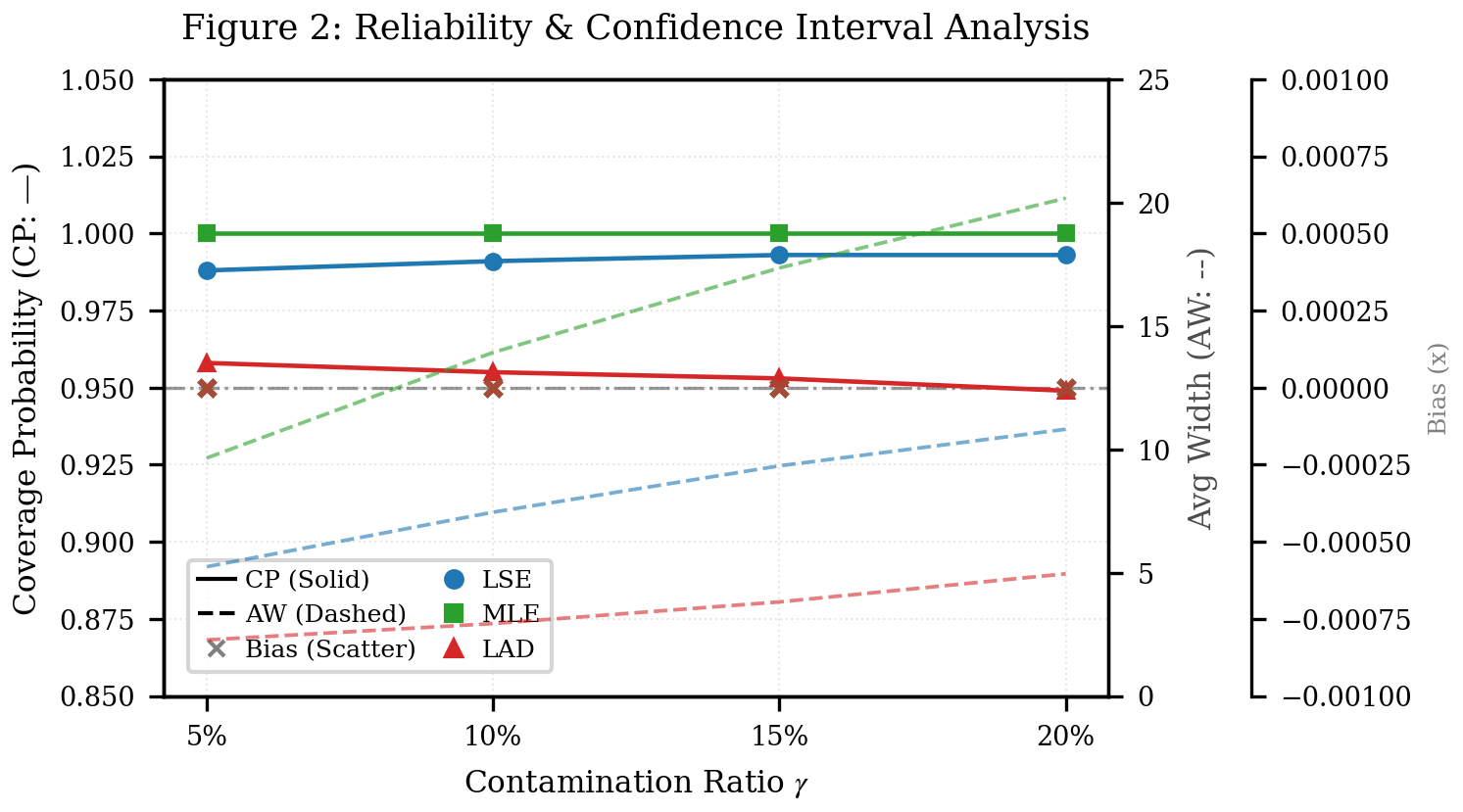}
	\end{minipage}
	\caption{Comparison of the three estimation methods under contaminated noise levels.}\label{fig2}
\end{figure}

In Fig. \ref{fig2}, as the contamination proportion $\gamma$ increases from $5\%$ to $20\%$, the estimation errors of all three methods exhibit an increasing trend due to the growing presence of outliers. However, their robustness characteristics differ substantially. The  LAD estimator demonstrates the greatest resistance to data contamination. At the highest contamination level of $20\%$,  $\mathrm{MSE}$ is only $1.52$, which is considerably lower than that of the  LSE ($3.26$) and the MLE ($6.62$), which is highly susceptible to contamination. These results provide empirical evidence that the absolute-deviation criterion is effective in mitigating the adverse effects of heavy-tailed disturbances and outliers.

Regarding interval estimation, the maximum likelihood estimator maintains extremely high coverage probabilities at the expense of substantially inflated confidence intervals, resulting in a marked loss of inferential precision. For instance, under the highest contamination level, the average interval width $\mathrm{AW}$ reaches $20.18$. The interval width of the least squares estimator also increases noticeably as the contamination level rises, indicating that its squared-loss mechanism is particularly sensitive to outliers. In contrast, the least absolute deviation estimator exhibits the slowest growth in interval width, with a maximum $\mathrm{AW}$ of only $4.96$, while its coverage probability $\mathrm{CP}$ consistently remains close to the nominal level of $0.95$.

Overall, all three estimators remain approximately unbiased from a theoretical perspective. The deterioration in performance is primarily attributable to the variance inflation induced by outliers rather than to systematic bias. When analyzing contaminated or heavy-tailed data, the least absolute deviation estimator, owing to its inherent robustness, achieves the most favorable balance between point-estimation accuracy and the reliability of interval inference.
\end{example}
\begin{example} (Independence Analysis under Block-Effect Fluctuations)\label{exam3} 
Given $\sigma=1$ and unchanged treatment effect $\tau$, two levels of block effects are considered: (i) Low-variation group: $a_{\text{low}}=[-0.5,-0.25,0,0.25,0.5]$, $b_{\text{low}}=[-0.2,-0.1,0,0.1,0.2]$; (ii) High-variation group: $a_{\text{high}}=[-5,-2.5,0,2.5,5]$, $b_{\text{high}}=[-4,-2,0,2,4]$.

\begin{figure}[htbp]
	\centering
	\begin{minipage}{0.48\textwidth}
		\centering
		\includegraphics[width=\textwidth]{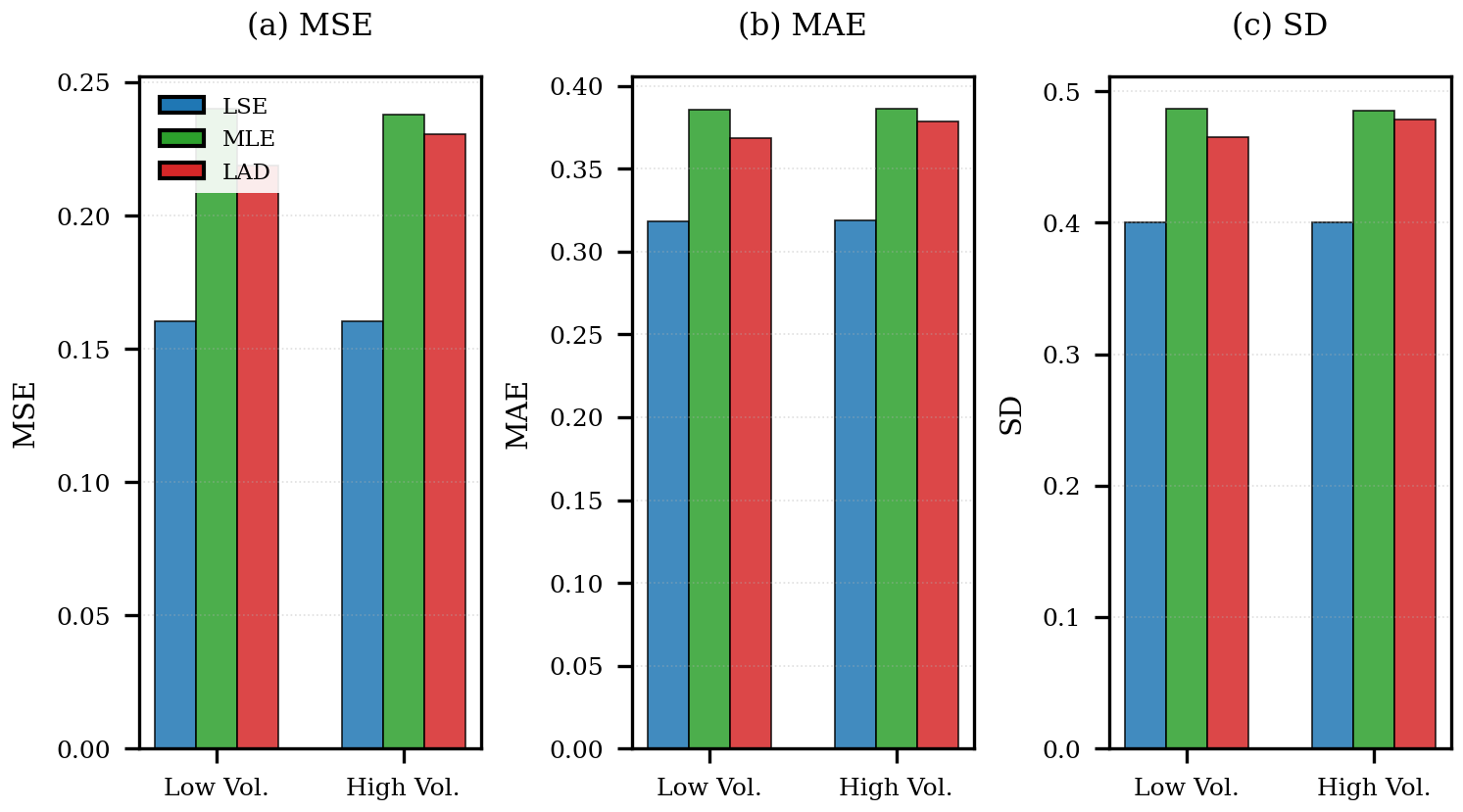}
	\end{minipage}
	\hfill
	\begin{minipage}{0.48\textwidth}
		\centering
		\includegraphics[width=\textwidth]{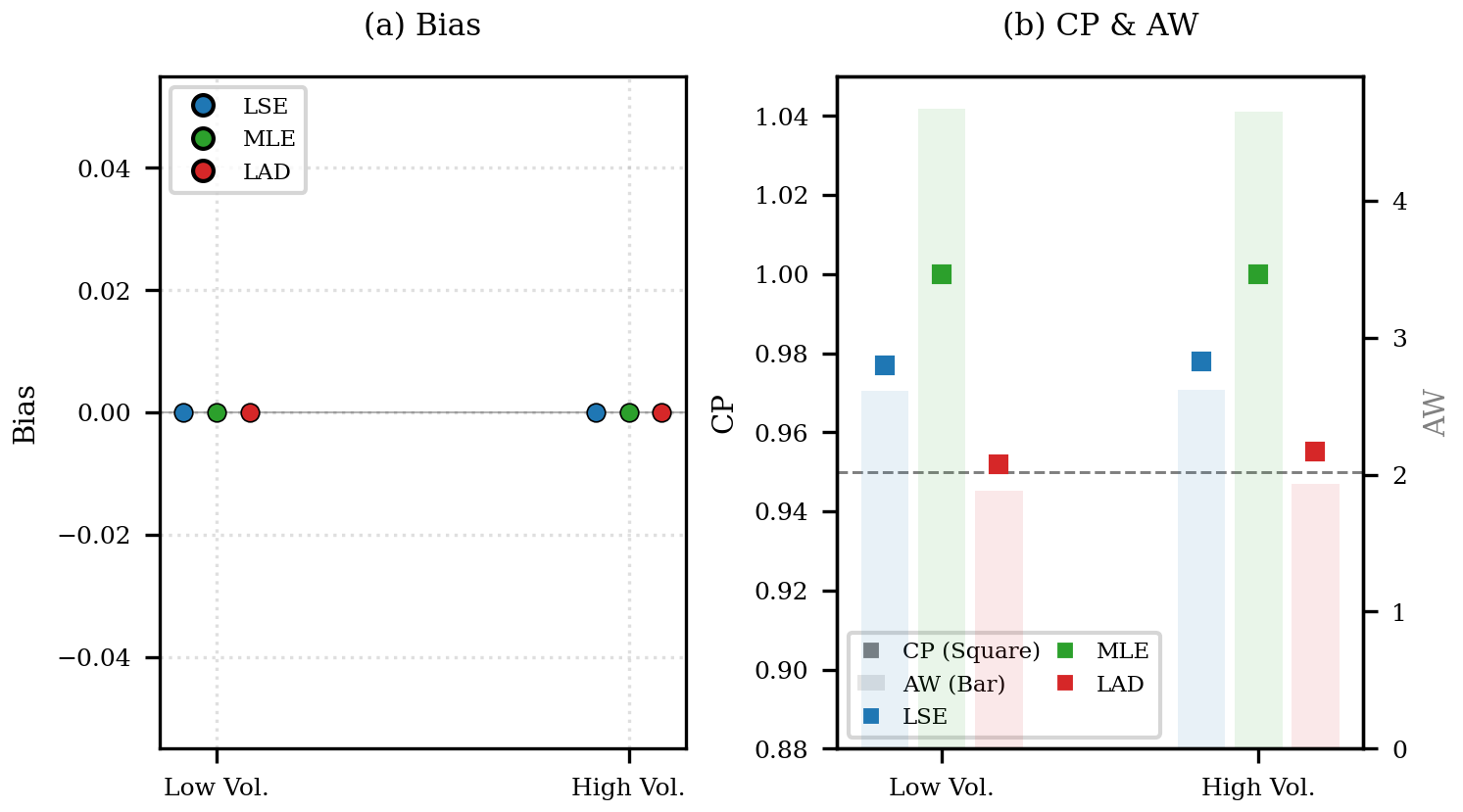}
	\end{minipage}
\end{figure}

Under the current background-variation scenario, characterized by transitions between low- and high-volatility conditions, all three estimation methods remain strictly unbiased, with the estimation bias consistently equal to zero. Moreover, each method exhibits high stability across changes in overall volatility.
Among them, the LSE demonstrates the strongest capability in data fitting and structural effect separation. $\mathrm{MSE}$ remains consistently at $0.16$, the lowest among all competing methods, across different volatility levels. In addition, $\mathrm{AW}$ is stably maintained at approximately $2.61$. These results indicate that LSE can effectively filter out background fluctuations and is therefore the preferred choice when highly accurate point estimation is the primary objective.
The LAD, in contrast, achieves the best performance from a statistical inference perspective. Its coverage probability $\mathrm{CP}$ remains consistently close to the nominal level of $0.95$, while its average interval width is the smallest among all methods, remaining below $2.0$. By attaining reliable coverage with the most compact confidence intervals, LAD demonstrates superior efficiency and effectiveness in interval estimation.
By comparison, the MLE appears overly conservative and relatively inefficient in this setting. To maintain complete coverage ($\mathrm{CP}=1.0$), it produces substantially inflated confidence intervals, with an average interval width of approximately $4.66$. At the same time, it yields the largest point-estimation error among the three methods.

In summary, if the primary concern is the accurate recovery of underlying treatment and block effects, LSE provides the most favorable performance. However, when the emphasis is on obtaining compact confidence intervals while maintaining coverage close to the nominal confidence level, LAD is the optimal choice.
\end{example}

The LSE, owing to its linear structural properties, achieves the highest statistical efficiency in handling and compensating for structured systematic variations, such as block effects. As a result, it can produce more precise point estimates and more efficient interval estimates in well-structured experimental settings.
The LAD, through its first-order absolute-deviation criterion, exhibits remarkable nonparametric robustness in the presence of unknown heterogeneous contamination and heavy-tailed disturbances. This robustness makes LAD particularly well suited for contaminated datasets, allowing it to maintain both the reliability and compactness of confidence intervals while preserving satisfactory estimation accuracy.
In contrast, the statistical performance of the MLE is heavily dependent on the validity of its underlying distributional assumptions. When confronted with complex nonparametric disturbances, such as structured noise, heterogeneous contamination, or heavy-tailed behavior, the robustness of its inferential framework can deteriorate substantially. Consequently, both point-estimation accuracy and interval-estimation efficiency may suffer significant degradation.
%%%%%%%%%%%%%%%%%%%%%%55

\subsection{Uncertain Estimators-based hypotheses}

Consider simulated $4 \times 4$ Latin square data. Three estimation methods, namely LSE, MLE, and LAD, are employed to validate the testing and inference procedures of the uncertain fixed-effects model. An experimental design with pronounced heteroscedastic characteristics is constructed. The observations are specified as
\[
y_{lij} = \mu + a_i + b_j + \tau_{l} + \epsilon_{lij},
\]
where $\mu = 50.0$, the row effects are $a = (0.5, -0.3, 0.2, -0.4)$, the column effects are $b = (-0.2, 0.4, -0.3, 0.1)$, and the treatment effects are $\tau = (0.3, -0.5, 0.4, -0.2)$. To simulate discrepancies in homogeneity tests caused by local strong fluctuations, the uncertain error term $\epsilon_{i,j}$ follows a heteroscedastic distribution, where the variance $\mathrm{Var}(\epsilon_{i,j})$ depends on the treatment level $l$: for $l \in \{0,1,2\}$, $\mathrm{Var}(\epsilon_{i,j}) = 0.0225$, while for $l=3$, the variance is set to $72.25$.

\begin{table}[htbp]
	\centering
	\caption{Simulated $4 \times 4$ Latin square data.}
	\label{tab1}
	\begin{tabular*}{\linewidth}{@{\extracolsep{\fill}}ccccc}
		\toprule
		\multirow{2}{*}{Row block} & \multicolumn{4}{c}{Column block} \\
		& $b_1$ & $b_2$ & $b_3$ & $b_4$ \\
		\midrule
		$a_1$ & 51.25 & 48.75 & 50.45 & 58.35 \\
		$a_2$ & 49.65 & 52.15 & 41.25 & 50.15 \\
		$a_3$ & 48.85 & 59.45 & 51.25 & 49.35 \\
		$a_4$ & 40.15 & 51.55 & 49.75 & 52.45 \\
		\bottomrule
	\end{tabular*}
\end{table}
To analyze the effects of row, column, and treatment factors on the observed responses, the following uncertain model is constructed:
\[
y_{lij} = \mu + a_i + b_j + \tau_{l} + \varepsilon_{lij}, \quad \varepsilon_{lij} \sim \mathcal{N}(0,\sigma),
\]
with constraints $\sum a_i = 0$, $\sum b_j = 0$, and $\sum \tau_l = 0$. The significance level is set to $\alpha = 0.05$, corresponding to a coverage probability of $\gamma = 0.95$. Parameter estimation is first performed, and the results are summarized in Table~\ref{tab:est}, where $\hat{c}$, $\tilde{c}$, and $\breve{c}$ denote the LSE, MLE, and LAD estimators, respectively. 

The adjusted residuals $\overset{*}e_{lij}$ are extracted from the model and projected onto three orthogonal dimensions: row ($R$), column ($C$), and treatment ($T$). The hypothesis $H_0^{(d)}: \sigma_1^{(d)} = \cdots = \sigma_k^{(d)}$ is tested accordingly.
At $\alpha = 0.05$, the quantile coefficient is computed as
$
K_\gamma = \frac{\sqrt{3}}{\pi}\ln\left(\frac{1.95}{0.05}\right) \approx 2.0198$.
The anomaly deviation proportion $\Pi_{s|r}^{(d)}$ between any two groups is calculated using Equation (12). Based on the rejection region $W^{(d)}$ defined in Equation (16), the null hypothesis of homogeneity is rejected if $\max \Pi_{s|r}^{(d)} \geq 0.05$. The test results and component standard deviation distributions are reported in Table~\ref{tab:est}.

The empirical results indicate that heteroscedasticity is detected under both LSE and LAD, whereas MLE assumes homoscedasticity. Subsequently, treatment effect inference is conducted separately under the three estimation methods.

\begin{table}[htbp]
	\centering
	\caption{Results of parameter estimation and homogeneity test ($\alpha=0.05$).}
	\label{tab:est}
	\renewcommand{\arraystretch}{1.3} 
	\small
	\begin{tabular}{llccc}
		\toprule
		 \multirow{2}{*}{Category} & \multirow{2}{*}{ Statistic / Dimension}  & \multicolumn{3}{c}{Estimation and Testing Methods} \\
		\cmidrule(lr){3-5}
		&& LSE ($\hat{c}$) & MLE ($\tilde{c}$) & LAD ($\breve{c}$) \\
		\midrule
		
		% --- Part I: Parameter Estimation ---
		\multirow{5}{*}{\begin{tabular}[c]{@{}l@{}}Parameter\\ Estimation\end{tabular}} 
		& Overall mean $\mu$ & 50.30 & 49.84 & 51.84 \\
		& Treatment effects $\tau$ & $[0.75, -0.93, 0.68, -0.50]$ & $[0.96, -0.47, -0.44, -0.04]$ & $[-1.61, -3.34, -1.56, 6.51]$ \\
		& Row effects $a$ & $[1.90, -2, 1.93, -1.83]$ & $[0, -0.18, 0.85, -0.68]$ & $[-0.46, -0.54, 0.39, 0.61]$ \\
		& Column effects $b$ & $[-2.83, 2.68, -2.13, 2.28]$ & $[-4.26, 4.09, -3.66, 3.84]$ & $[-1.81, 0.71, 0.64, 0.46]$ \\
		& Global scale $\sigma$ & 5.55 & 5.55 & 10.10 \\
		
		\midrule
		
		% --- Part II: Homogeneity Tests ---
		\multirow{9}{*}{\begin{tabular}[c]{@{}l@{}}Homogeneity\\ Tests\end{tabular}} 
		& Rows & & & \\
		& Standard deviations $\sigma_r$ & $[3.45, 3.44, 3.35, 3.35]$ & $[5.55, 5.55, 5.55, 5.55]$ & $[1.65, 8.82, 0, 8.52]$ \\
		& Test statistic/Conclusion & $0$ (Accept) & $0$ (Accept) & $0$ (Reject) \\
		\cmidrule(lr){2-5}
		
		& Columns & & & \\
		& Standard deviations $\sigma_c$ & $[3.67, 3.64, 3.10, 3.14]$ & $[5.55, 5.55, 5.55, 5.55]$ & $[8.83, 0.85, 8.60, 0.55]$ \\
		& Test statistic/Conclusion & $0$ (Accept) & $0$ (Accept) & $0$ (Reject) \\
		\cmidrule(lr){2-5}
		
		& Treatments & & & \\
		&Standard deviations $\sigma_t$ & $[0.86, 4.73, 0.85, 4.72]$ & $[5.55, 5.55, 5.55, 5.55]$ & $[1.65, 1.75, 1.01, 12.09]$ \\
		& Test statistic/Conclusion & $1$ (Reject) & $0$ (Accept) & $0.5$ (Reject) \\
		
		\bottomrule
	\end{tabular}
\end{table}

\begin{example}(Treatment Effect Testing under Maximum Likelihood Estimation)
Based on the previous analysis, the  MLE approach accepts the homogeneity hypothesis
$H_0^{(T)} \cap H_0^{(R)} \cap H_0^{(C)}$
across the treatment, row, and column orthogonal dimensions. To further examine whether the global standard deviation $\tilde{\sigma}$ equals a fixed constant $\tilde{\sigma_0}$, we consider the following hypotheses:
\[
H_0^{\sigma}: \tilde{\sigma} = \tilde{\sigma_0} = 5.55 
\quad \text{vs} \quad 
H_1^{\sigma}: \tilde{\sigma} \neq 5.55.
\]
At a significance level of $\alpha=0.05$, the quantile coefficient is computed as
$K_\gamma = \frac{\sqrt{3}}{\pi}\ln\frac{2-\alpha}{\alpha} \approx 2.0198$.
The corresponding decision threshold is therefore
$
\Phi^{-1}(1-\alpha/2; \tilde{\sigma_0}) = \sigma_0 K_\gamma = 11.21.
$
Let $\mathbf{E}$ denote the collection of $N=16$ residual observations. According to the rejection-region construction rule, all residuals fall within the acceptance region $[-11.21, 11.21]$, with the number of abnormal observations equal to $M=0$, i.e., $\mathbf{E} \notin W^{\sigma}$. At the 0.05 significance level, $H_0^{\sigma}$ cannot be rejected, indicating that the Latin square model satisfies homoscedasticity and that the common standard deviation equals $\tilde{\sigma_0} = 5.55$.

Under the validity of this assumption, we further investigate whether treatment effects are statistically significant. The following homogeneity hypothesis is considered:
\[
H_0^{t}: \tau_1 = \tau_2 = \tau_3 = \tau_4 
\quad \text{vs} \quad 
H_1^{t}: \tau_1, \tau_2, \tau_3, \tau_4 \text{ are not all equal}.
\]
According to the rejection rule in Equation (15), In all pairwise comparisons, the anomaly deviation proportion $\Pi_{s|r}$ equals 0. Since $	\Pi_{s|r}^{(t1)} = 0 < 0.05,$, we fail to reject $H_0^{\tau}$, indicating treatment homogeneity. The distributional characteristics and interval coverage of each group are summarized in Table~\ref{tab:homo}.

After passing the homogeneity test, we further examine whether the overall treatment effect differs significantly from zero:
\[
H_0^{tc}: \tau = 0 
\quad \text{vs} \quad 
H_1^{tc}: \tau \neq 0.
\]
The results show that all 16 adjusted observations fall within the null decision boundary $\pm 11.21$, with anomaly proportion $\Pi^{(tc1)} = 0$. At the 0.05 significance level, we fail to reject $H_0^{t c}$, indicating that the treatment effect is not statistically significant.
\end{example}
%%%%%%%%%%%%%%%%%%%%%%%%%%%%%%%%%%%%%%%%%%%%%%%%%%%%
\begin{example}(Treatment Effect Testing under Least Squares Estimation)
According to the estimated scale parameters, the LSE method exhibits significant heteroscedasticity. Therefore, a robust inference procedure based on local standard deviations $\hat{\sigma}_r$ (Case ii) is adopted. We consider the homogeneity hypothesis:
\[
H_0^{t}: \tau_1 = \tau_2 = \tau_3 = \tau_4 
\quad \text{vs} \quad 
H_1^{t}: \text{not all equal}.
\]
At the significance level $\alpha=0.05$, the acceptance intervals
are summarized in Table~\ref{tab:homo}.
The maximum anomaly deviation proportion is $\Pi_{s|r}^{t2} = 1 > 0.05$, thus, $H_0^{t}$ is rejected, indicating significant non-homogeneity among treatment levels under LSE.
\end{example}
%%%%%%%%%%%%%%%%%%%%%%%%%%%%%%%%%%%%%%%%%%%%%%%%%%
\begin{example}(Treatment Effect Testing under Least Absolute Deviations)
Since the LAD method also exhibits heteroscedasticity, the same robust inference scheme (Case ii) is applied.
Under the homogeneity hypothesis $H_0^{t}$ and $\alpha=0.05$, the acceptance intervals are reported in Table~\ref{tab:homo}.
The results show that the maximum anomaly deviation proportion is $\Pi_{s|r}^{(t1)} = 1 > 0.05$, thus $H_0^{t}$ is rejected, indicating that the treatment levels are not homogeneous under the LAD estimator as well.
\end{example}
\begin{table}[htbp]
	\centering
	\caption{Acceptance intervals for treatment effect homogeneity tests}
	\label{tab:homo}
	\renewcommand{\arraystretch}{1.5}
	\small
	\begin{tabular*}{\linewidth}{@{\extracolsep{\fill}}cccc}
		\toprule
		Level & MLE Acceptance Interval & LSE Acceptance Interval & LAD Acceptance Interval \\
		\midrule
		$T_1$ & $[-10.26, 12.17]$ & $[-0.98, 2.48]$  & $[-4.95, 1.72]$ \\
		$T_2$ & $[-11.68, 10.74]$ & $[-10.48, 8.63]$ & $[-6.87, 0.20]$ \\
		$T_3$ & $[-11.66, 10.77]$ & $[-1.04, 2.39]$  & $[-3.61, 0.48]$ \\
		$T_4$ & $[-11.26, 11.17]$ & $[-10.04, 9.04]$ & $[-17.91, 30.94]$ \\
		\bottomrule
	\end{tabular*}
\end{table}

In summary, the three methods yield markedly different inferential conclusions. Due to its overly flexible estimation of the global scale parameter, MLE fails to detect variance heterogeneity across dimensions, resulting in excessively wide acceptance regions and ultimately a conservative conclusion of homogeneity and insignificance. In contrast, both LSE and LAD are able to sensitively capture heteroscedasticity in the treatment dimension and consistently reject the homogeneity hypothesis under a robust testing framework, indicating significant treatment effects.
Overall, compared with MLE,  both LSE and LAD demonstrate higher power and improved signal detection.
%%%%%%%%%%%%%
\subsection{Empirical Analysis}
The dataset used in this study was obtained from Kamfanin Bobi Public Secondary School in Niger State, Nigeria. During the study, teachers employing different instructional methods across classes were observed, and students’ academic performance was evaluated \cite{28}.

\begin{table}[htbp]
	\centering
	\caption{Experimental results of treatment--block combinations in Latin square design.}
	\begin{tabular*}{\linewidth}{@{\extracolsep{\fill}}lcccccc@{}}
		\toprule
		\multirow{2}{*}{\makecell{Column block\\(Grade level)}} & \multicolumn{6}{c}{Row block (Teaching ability of teachers)} \\
		\cmidrule{2-7}
		& Mohammed & Usman & Marcus & Ahmed & Sanusi & Mustafa \\
		\midrule
		Junior Year 1 & $A=35$ & $D=17$ & $C=16$ & $E=18$ & $B=44$ & $F=17$ \\
		Junior Year 2 & $B=11$ & $A=42$ & $E=14$ & $C=21$ & $F=40$ & $D=39$ \\
		Junior Year 3 & $C=29$ & $E=28$ & $D=45$ & $F=34$ & $A=19$ & $B=49$ \\
		Senior Year 1 & $D=23$ & $C=17$ & $F=36$ & $B=24$ & $E=41$ & $A=16$ \\
		Senior Year 2 & $F=21$ & $B=45$ & $A=31$ & $D=14$ & $C=31$ & $E=35$ \\
		Senior Year 3 & $E=19$ & $F=28$ & $B=13$ & $A=25$ & $D=24$ & $C=19$ \\
		\bottomrule
	\end{tabular*}
\end{table}
An uncertain Latin square model is constructed as
\[
y_{lij} = \mu + a_i + b_j + \tau_{l} + \varepsilon_{lij}, 
\qquad \varepsilon_{lij} \sim \mathcal{N}(0,\sigma),
\]
subject to the constraints $\sum a_i = 0$, $\sum b_j = 0$, and $\sum \tau_l = 0$. Parameter estimation is first conducted, and the results of the three methods are reported in~Table \ref{tab5}.
The adjusted residuals $\overset{*}e_{lij}$ are then extracted and projected onto three orthogonal dimensions: row ($R$), column ($C$), and treatment ($T$). The hypothesis
\[
H_0^{(d)}: \sigma_1^{(d)} = \cdots = \sigma_6^{(d)}
\]
is tested for each dimension $d \in \{T, R, C\}$.  With significance level $\alpha = 0.05$, the quantile coefficient is $K_\gamma \approx 2.0198$. For each dimension $d$, the anomaly deviation proportion matrix $\Pi_{s|r}^{(d)}$ is computed according to Equation (12) to assess heteroscedasticity. Table~\ref{tab5}  summarizes the testing results for LSE, MLE, and LAD across the three dimensions.

\begin{table}[htbp]
	\centering
	\caption{Results of parameter estimation, local standard deviations, and homogeneity tests ($\alpha=0.05$).}
	\label{tab5}
	\renewcommand{\arraystretch}{1.35}
	\footnotesize
	\begin{tabular*}{\linewidth}{@{\extracolsep{\fill}}lcccc@{}}
		\toprule
		Category & Item / Dimension & LSE ($\hat{c}$) & MLE ($\tilde{c}$) & LAD ($\breve{c}$) \\
		\midrule
		\multirow{5}{*}{\begin{tabular}[c]{@{}l@{}}Panel A\\Estimate\end{tabular}} 
		& Overall mean $\mu$ & 27.22 & 24.92 & 28.33 \\
		\cmidrule{2-5}
		& Treatment effects $\tau$ & 
		\makecell[c]{[0.78, 3.78, -5.06] \\ {[-0.22, -1.39, 2.11]}} & 
		\makecell[c]{[1.11, 4.53, -9.14] \\ {[0.36, 1.03, 2.11]}} & 
		\makecell[c]{[11.17, -0.83, -6.17] \\ {[-3.50, -1.17, 0.50]}} \\
		
		& Row effects $a$ & 
		\makecell[c]{[-2.72, 0.61, 6.78] \\ {[-1.06, 2.28, -5.89]}} & 
		\makecell[c]{[0.90, -0.68, 5.15] \\ {[1.40, 4.40, -11.18]}} & 
		\makecell[c]{[-0.56, -0.56, 10.78] \\ {[2.11, -2.89, -8.89]}} \\
		
		& Column effects $b$ & 
		\makecell[c]{[-4.22, 2.28, -1.39] \\ {[-4.56, 5.94, 1.94]}} & 
		\makecell[c]{[-4.85, 3.74, 1.65] \\ {[-2.76, 0.74, 1.49]}} & 
		\makecell[c]{[-3.94, -7.28, -5.61] \\ {[-5.61, 11.72, 10.72]}} \\
		
		& Global std $\sigma$ & 11.84 & 15.22 & 16.50 \\
		
		\midrule
		
		% --- Panel B: Scale Parameters and Homogeneity Tests ---
		\multirow{9}{*}{\begin{tabular}[c]{@{}l@{}}Panel B\\Scale/Test\end{tabular}} 
		
		& Treatment Effects (T) & & & \\
		& Local standard deviations $sd_\tau$ & 
		\makecell[c]{[13.01, 10.46, 3.47] \\ {[8.81, 7.20, 6.90]}} & 
		\makecell[c]{[15.22, 15.22, 15.22] \\ {[15.22, 15.22, 15.22]}} & 
		\makecell[c]{[23.37, 12.50, 2.89] \\ {[6.58, 3.81, 11.79]}} \\
		
		& Test conclusion & Reject & Accept & Reject \\
		\cmidrule{2-5}
		
		& Row Blocks (R) & & & \\
		& Local standard deviations $sd_a$ & 
		\makecell[c]{[9.30, 10.23, 11.47] \\ {[8.16, 6.76, 5.72]}} & 
		\makecell[c]{[15.22, 15.22, 15.22] \\ {[15.22, 15.22, 15.22]}} & 
		\makecell[c]{[9.32, 7.54, 18.62] \\ {[15.46, 11.34, 7.02]}} \\
		& Test conclusion & Reject & Accept & Reject  \\
		\cmidrule{2-5}
		& Column Blocks (C) & & & \\
		& Local standard deviations $sd_b$ & 
		\makecell[c]{[9.44, 8.14, 9.02] \\ {[5.57, 10.80, 9.13]}} & 
		\makecell[c]{[15.22, 15.22, 15.22] \\ {[15.22, 15.22, 15.22]}} & 
		\makecell[c]{[5.27, 13.63, 8.04] \\ {[2.56, 17.75, 17.54]}} \\
		& Test conclusion & Reject & Accept & Reject  \\
		\bottomrule
	\end{tabular*}
\end{table}

\textit{Significance Test of Teaching Methods.} This study aims to examine whether teaching methods (i.e., treatment effects) have a statistically significant impact on students’ academic performance. Since students’ scores do not exhibit frequency stability, and their generating distribution may deviate from future empirical frequencies, the observed scores are treated as uncertain variables rather than random variables.
We conduct the following hypothesis test using three estimation methods:
\[
H_0: \text{Teaching methods have no significant effect }
\quad \text{vs} \quad
H_1: \text{Teaching methods have a significant effect}.
\]
Next, we study the homogeneity and the common test of treatment effects across three estimation methods. 

%%%%%%%%%%%%%%%%%%%%%%%%%%%%%%%%%%%%%%%%%%%%
Case (i): Treatment Effect Testing under Maximum Likelihood Estimation.
%%%%%%%%%%%%%%%%%%%%%%%%%%%%%%%%%%%%%%%%%%%%
Based on the previous analysis, the  MLE approach accepts the homogeneity hypothesis
$H_0^{(T)} \cap H_0^{(R)} \cap H_0^{(C)}$
across treatment, row, and column dimensions. To further test $\tilde{\sigma}=\tilde{\sigma_0}$, consider the following hypotheses:
\[
H_0^{\sigma}: \tilde{\sigma} = \tilde{\sigma_0} = 15.22 
\quad \text{vs} \quad 
H_1^{\sigma}: \tilde{\sigma} \neq 15.22.
\]
At a significance level of $\alpha=0.05$, the quantile coefficient is computed as
$K_\gamma = \frac{\sqrt{3}}{\pi}\ln\frac{2-\alpha}{\alpha} \approx 2.0198$,
yielding the decision threshold
$
\Phi^{-1}(1-\alpha/2;\tilde{\sigma_0}) = \tilde{\sigma_0} K_\gamma = 30.74.
$
Let $\mathbf{E}$ denote the collection of all $N=36$ residual observations. According to the rejection region construction rule, the null hypothesis $H_0^{\sigma}$ is rejected if at least $\alpha N = 1.8$ (i.e., at least 2) observations satisfy $|\tilde{e}_m| > 30.74$.
The results show that all residuals fall within the acceptance region $[-30.74, 30.74]$. At the 0.05 significance level, $H_0^{\sigma}$ cannot be rejected, indicating that the model satisfies homoscedasticity, and the common standard deviation equals the fixed constant $\tilde{\sigma}_0 = 15.22$.

Under this assumption, we further investigate whether the treatment factor has a significant effect on the response variable. The following homogeneity hypothesis is considered:
\[
H_0^{\tau}: \tau_A = \tau_B = \tau_C = \tau_D = \tau_E = \tau_F 
\quad \text{vs} \quad 
H_1^{\tau}: \text{not all treatment effects are equal}.
\]
According to the rejection rule in Equation (12), the anomaly deviation proportion $\Pi_{s|r}$ equals 0 for all pairwise comparisons (with maximum $\max \Pi_{l|r} = 0$). Since $\Pi_{s|r} < \alpha$, we fail to reject $H_0^{t}$. The results indicate treatment homogeneity. The distributional characteristics and interval coverage for each group are summarized in Table~\ref{tab6}.
After passing the homogeneity test for treatment effects, we further examine whether the overall treatment effect is significantly different from zero:
\[
H_0^{\tau c}: \tau = 0 
\quad \text{vs} \quad 
H_1^{\tau c}: \tau \neq 0.
\]

The results show that all 36 adjusted observations fall within the zero-level decision boundary $\pm 30.74$, with anomaly proportion $\Pi^{(tc1)} = 0$. At the 0.05 significance level, we fail to reject $H_0^{tc}$, indicating that the treatment effect is not statistically significant.

%%%%%%%%%%%
Case (ii):  Treatment Effect Testing under the Least Squares Method.
%%%%%%%%%
Based on the estimated scale parameters, the LSE method exhibits pronounced heteroscedasticity.
We consider the following homogeneity hypothesis:
\[
H_0^{\tau}: \tau_1 = \tau_2 = \tau_3 = \tau_4 = \tau_5 = \tau_6 
\quad \text{vs} \quad 
H_1^{\tau}: \text{not all treatment effects are equal}.
\]
At the significance level $\alpha=0.05$, the quantile coefficient is computed as $K_\gamma \approx 2.0198$.
The resulting acceptance intervals $AI(\cdot; \hat{\tau}_r, \hat{*}{\sigma}_r)$ are summarized in ~Table \ref{tab6}.
It is found that, across pairwise comparisons between different treatment levels, the maximum anomaly deviation proportion is $\max \Pi_{s|r}^{t2} = 1$.
Since $\max p_{s|r}^{t2} \geq 0.05$, the results fall into the rejection region $W^{t2}$.
Therefore, we reject the null hypothesis $H_0^t$, concluding that under the LSE framework, the treatment levels exhibit significant heterogeneity and the treatment effects differ significantly.
%%%%%%%%%%%%%%%%%%%%%%%%%%%%%%%%%%%%%%%%%%%%%%%%%%

Case (iii): Treatment Effect Testing under the Least Absolute Deviation Method.
%%%%%%%%%%%%%%%%%%%%%%%%%%%%%%%%%%%%%%%%%%%%%%%%%%%%
Since the LAD method also exhibits heteroscedasticity, we adopt the same robust inference procedure under $H_0^{t}$.
At the significance level $\alpha=0.05$, the quantile coefficient is $K_\gamma \approx 2.0198$.
The corresponding acceptance intervals obtained from the LAD method are reported in Table~\ref{tab6}.
The data validation shows that, across pairwise comparisons between treatment levels, the maximum anomaly deviation proportion is $\max \Pi_{s|r}^{t2} = 1$.
Since $\max\Pi_{s|r}^{t2} \geq 0.05$ (i.e., the results fall into the rejection region $W^{t2}$), we reject the null hypothesis $H_0^t$, concluding that under the LAD framework, the treatment levels also fail to satisfy homogeneity, and significant treatment differences exist.

\begin{table}[htbp]
	\centering
	\caption{Acceptance intervals for treatment effect homogeneity tests}\label{tab6}
	%\label{tab:omo}
	\renewcommand{\arraystretch}{1.5}
	\small
	\begin{tabular*}{\linewidth}{@{\extracolsep{\fill}}cccc@{}}
		\toprule
		Level & LSE (Ordinary Least Squares) & MLE (Maximum Likelihood) & LAD (Least Absolute Deviations) \\
		\midrule
		A & $[-25.50, 27.06]$ & $[-29.62, 31.84]$ & $[-36.03, 58.36]$ \\
		B & $[-17.35, 24.90]$ & $[-26.21, 35.26]$ & $[-26.09, 24.42]$ \\
		C & $[-12.05, 1.94]$  & $[-39.87, 21.59]$ & $[-12.00, -0.34]$ \\
		D & $[-18.01, 17.57]$ & $[-30.37, 31.09]$ & $[-16.79, 9.79]$ \\
		E & $[-15.93, 13.15]$ & $[-29.71, 31.76]$ & $[-8.85, 6.52]$ \\
		F & $[-11.82, 16.04]$ & $[-28.62, 32.84]$ & $[-23.31, 24.31]$ \\
		\bottomrule
	\end{tabular*}
\end{table}
%%%%%%%%%%%%%%%%%%%%%%%%%%%%%%%%%%%%%%%%%%%%%%%%%%%%
%%%%%%%%%%%%%%%%%%%%%%%%%%%%%%%%%%%%%%%%%%%%%%%%%%%%
After controlling for teacher and class blocking effects, both LSE and LAD estimators indicate that different teaching methods have a significant impact on students’ academic performance. This conclusion is consistent with the findings in the study \cite{29}, confirming the effectiveness of the proposed approach.

\begin{remark}
	Under heteroscedastic conditions, the MLE approach fails in inference due to scale inflation, whereas LSE and LAD successfully capture treatment significance by identifying local variability, demonstrating superior robustness and statistical efficiency.
\end{remark}

\section{Conclusion}\label{s6}
This paper mainly established an uncertain fixed-effects model and investigated the estimation and inference for Latin-square designs within the framework of uncertainty theory. For experimental data lacking frequency stability, three parameter estimation methods and their corresponding effect-testing strategies are studied, with the aim of providing a possible analytical approach for handling uncertain data in complex experimental environments. 
Some advances have been made in parameter estimation and hypothesis testing research. By introducing LSE, MLE, and LAD, along with their confidence intervals, this study aims to establish a comprehensive framework for parameter estimation. In addition, the proposed uncertain homogeneity test and common test provide a reference basis for assessing the significance of treatment effects. Numerical simulation results indicate that different estimation methods exhibit distinct advantages under different experimental scenarios:
\begin{enumerate}
	\item In scenarios with low data contamination and highly structured experimental settings, LSE demonstrates relatively high statistical efficiency. Its confidence intervals are comparatively narrower, which is advantageous for the refined separation of treatment effects.
	\item In extreme scenarios involving significant outliers or high contamination levels, LAD exhibits strong robustness. When the experimental error distribution is disturbed by outliers, its parameter estimates can still maintain satisfactory consistency.
	\item The inferential conclusions of MLE show a certain degree of sensitivity to the underlying distributional assumptions of the model. Therefore, practical applications require careful consideration alongside the characteristics of the observed data.
\end{enumerate}
Through numerical simulations and an empirical analysis of educational data, the feasibility of the proposed models is preliminarily validated. The results suggest that the proposed approach demonstrates some analytical effectiveness in handling non-ideal experimental data and may serve as a useful reference for experimental design in related fields.

Although this study presents an initial exploration of uncertain modeling in Latin square designs, the application of uncertainty theory to experimental design is still at an early stage of development, and several limitations remain. For example, the present work does not yet provide an in-depth discussion of model performance under missing-data conditions or in more complex experimental designs, such as Graeco-Latin square designs. Future research may further focus on extending the model structure and exploring parameter-update strategies in dynamic environments, with the aim of improving the general applicability of the proposed framework across broader domains.

\section*{Conflict of interest} 
The authors declare that they have no conflict of interest.

\section*{Data Availability }  
The data used in the motivating example are from Sule, et al. \cite{28}.

\section*{Funding}
The work was supported by the National Natural Science Foundation of China (12561047),  the Xinjiang Talent Development Fund (XJRC-2025-KJ-PY-KJLJ-108), and the 2025 Central Guidance for Local Science and Technology Development Fund (ZYYD2025ZY20).

%%%%%%%%%%%%%%%

\end{document}